\documentclass[12pt, draftclsnofoot, onecolumn]{IEEEtran}
\usepackage[T1]{fontenc}
\usepackage{cite}
\usepackage{graphicx,amsmath,amssymb,bm,cite,algorithm,algpseudocode,amsthm}

\newcommand{\h}{\mathbf{h}}

\newcommand{\n}{\mathbf{n}}

\newcommand{\s}{\mathbf{s}}

\newcommand{\w}{\mathbf{w}}

\newcommand{\y}{\mathbf{y}}

\newcommand{\0}{\mathbf{0}}

\newcommand{\A}{\mathbf{A}}
\newcommand{\B}{\mathbf{B}}
\newcommand{\C}{\mathbf{C}}
\newcommand{\D}{\mathbf{D}}
\newcommand{\E}{\mathbf{E}}
\newcommand{\F}{\mathbf{F}}
\newcommand{\G}{\mathbf{G}}
\renewcommand{\H}{\mathbf{H}}
\newcommand{\I}{\mathbf{I}}

\newcommand{\Q}{\mathbf{Q}}

\newcommand{\U}{\mathbf{U}}
\newcommand{\V}{\mathbf{V}}
\newcommand{\W}{\mathbf{W}}

\newcommand{\Real}{\mbox{$\mathbb{R}$}}
\newcommand{\Compl}{\mbox{$\mathbb{C}$}}

\newcommand{\tr}{\mathrm{tr}}

\begin{document}
\title{Low Complexity Full Duplex MIMO:\\ Novel Analog Cancellation Architectures and Transceiver Design}
\author{George~C.~Alexandropoulos,~\IEEEmembership{Senior Member,~IEEE,} and Melissa~Duarte,~\IEEEmembership{Member,~IEEE}\thanks{Part of this work has been presented in \textit{IEEE ICC}, Paris, France, 21--25 May 2017 \cite{GM_ICC_2017}.}\thanks{G.~C.~Alexandropoulos and M. Duarte are with the Mathematical and Algorithmic Sciences Lab, Paris Research Center, Huawei Technologies France SASU, 92100 Boulogne-Billancourt, France, e-mails: \{george.alexandropoulos, melissa.duarte\}@huawei.com.}}

\maketitle
\begin{abstract}
Incorporating full duplex operation in Multiple Input Multiple Output (MIMO) systems provides the potential of boosting throughput performance. However, the hardware complexity of the analog self-interference canceller in emerging full duplex MIMO designs mostly scales with the number of transmit and receive antennas, thus exploiting the benefits of analog cancellation becomes impractical for full duplex MIMO transceivers, even for moderate number of antennas. In this paper, we present two novel architectures for the analog canceller comprising of reduced number of cancellation elements, compared to the state of the art, and simple multiplexers for efficient signal routing among the transmit and receive radio frequency chains. One architecture is based on analog taps (tap refers to a line of fixed delay, variable phase shifter, and attenuator) and the other on AUXiliary (AUX) Transmitters (TXs) that locally generate the cancellation signal. In contrast to the available analog cancellation architectures, the values for each tap or each AUX TX and the configuration of the multiplexers are jointly designed with the digital transmit and receive beamforming filters according to certain performance objectives. Focusing on a narrowband flat fading channel model as an example, we present a general optimization
framework for the joint design of analog self-interference cancellation and digital beamforming. We also detail the sum rate optimization objective together with its derived solution for the latter architectural components. Representative computer simulation results demonstrate the superiority both in terms of hardware complexity and achievable performance of the proposed low complexity full duplex MIMO schemes over the lately available ones.
\end{abstract}

\IEEEpeerreviewmaketitle

\section{Introduction}
In band full duplex, also known shortly as Full Duplex (FD), is a candidate technology for fifth Generation (5G) wireless systems because of the potential spectral efficiency gains that can be achieved through simultaneous uplink and downlink communication within the entire frequency band \cite{Sab14_all, Am45G14}. An FD radio can transmit and receive at the same time and same frequency resource unit, consequently, it can double the spectral efficiency achieved by a half duplex radio. Current wireless systems exploit Multiple Input Multiple Output (MIMO) communication, where increasing the number of transmit and receive antennas can increase the spatial Degrees of Freedom (DoF), hence boosting spectral efficiency. Combining FD with MIMO communication can provide further spectral efficiency gains \cite{Rii11_all, Nguyen2013_all, Bha14, SofNull_2016, Tam2016_all, Italo_EW_2016_all, GA2016_all}. Thus, enabling FD MIMO technology, for small to large antenna array systems, is of high interest in order to achieve the demanding throughput requirements of 5G wireless communication systems \cite{Shafi_JSAC_5G}. 

An FD radio suffers from Self Interference (SI), which is the signal transmitted by the FD radio Transmitter (TX) that leaks to the FD radio Receiver (RX). At the RX of the FD radio, the power of the SI signal can be many times stronger than the power of the received signal of interest (which is transmitted from another radio). Consequently, SI can severely degrade the reception of the signal of interest, and thus SI mitigation is required in order to maximize the spectral efficiency gain of the FD operation. As the number of antennas increases, mitigating SI becomes more challenging, since more antennas naturally result in more SI components. For the case of a Single Input Single Output (SISO) FD node, it has been demonstrated \cite{Bha13_all,Dua14_all} that significant SI mitigation can be achieved via a combination of analog and digital cancellation techniques, where an estimate of the received SI is subtracted from the received signal (which is the sum of the SI signal and signal of interest). A straightforward extension of SI mitigation solutions used in SISO FD to the case of MIMO FD can be envisioned. However, the hardware resources required for analog SI cancellation become the main bottleneck, since they scale with the number of antenna elements. Specifically, for the two most widely considered analog canceller solutions, which are: \textit{i}) the architecture based on taps (a tap consists of analog components that implement delay, phase shift, and attenuation) \cite{Bha13_all,Kol16_all}; and \textit{ii}) the architecture based on AUXiliary (AUX) TX Radio Frequency (RF) chains (a AUX TX RF chain generates an analog cancellation signal from an input digital reference signal) \cite{Dua14_all,Hub15}, the hardware requirements in MIMO scenarios are as follows. For the case where the analog canceller is based on multiple taps, an extension to MIMO requires at least $M_kN_k$ taps with $M_k$ and $N_k$ denoting the number of RX and TX antennas, respectively, at a FD MIMO node $k$. For the case where the analog canceller is based on multiple AUX TX RF chains, an extension to MIMO requires at least $M_k$ AUX TXs. Consequently, depending on the number of TX and RX antennas at the FD MIMO node, the extension of SISO analog canceller solutions to the MIMO case may be prohibitively complex. Thus, recent works have proposed only digital SI mitigation for FD MIMO \cite{Rii11_all,SofNull_2016}. These approaches exploit the availability of multiple antennas at the FD node in order to provide SI mitigation via digital BeamForming (BF); such an approach is known as spatial suppression. However, as has been pointed out, spatial suppression approaches often result in lower rates for both the outgoing and incoming signals of interest, since some of the available spatial DoF are solely devoted for mitigating SI.

In this paper we propose two novel architectures for analog SI cancellation and a novel optimization framework for jointly designing the analog canceller and the TX/RX digital BF parameters. The first new architecture for analog cancellation consists of multi-tap hardware, where the number of taps does not increase with the number of TX or RX antenna elements. The second new architecture includes AUX TX RF chains whose number does not depend on the number of TX or RX antennas. The number of taps in the one architecture and that of AUX TXs in the other can be chosen offline as a function of size constraints, cost per tap and cost per AUX TX RF chain, or other constraints on the analog canceller hardware. Both simplified analog canceller architectures are enabled via the use of MUltipleXers(MUXs) and DEMUltipleXers(DEMUXs), which allow flexible connectivity between the taps or AUX TXs and the transceiver antennas. The settings of taps or AUX TXs and the configurations of MUXs/DEMUXs is computed via our proposed optimization framework. The flexible signal routing via MUXs/DEMUXs enables the use of reduced taps or AUX TXs in an optimized way, since either of the latter will be used between the subset of TX and RX antennas where they are mostly beneficial. The digital beamformer and analog canceller parameters are thus designed by taking into account each others capabilities, hence the burden of SI mitigation is split between digital BF and analog cancellation. We note that the related work \cite{Rii11_all} has considered joint design of digital BF and analog cancellation, however these and related solutions \cite{Atz16_all,Zha12_all} assume underlying analog canceller hardware as in \cite{Bha13_all,Bha14,Kol16_all,Hub15,Dua14_all}, which scales with the number of transceiver antennas. For the JointNull solution recently proposed in \cite{Gow_18_all}, although the number of analog cancellers does not necessarily scale with the number of antennas, the underlying architecture of the canceller (i$.$e$.$, number of taps or AUX TXs) is not taken into account in the BF design.  As our simulation results will show, our proposed analog canceller architecture together with our novel joint design of analog cancellation and TX/RX digital BF is capable of achieving higher rates with less hardware compared to State-of-the-Art (SotA) FD MIMO solutions. This paper's contributions can be summarized as follows. 
\begin{itemize}
\item We present two novel analog SI canceller architectures, one based on multiple taps and another one consisting of multiple AUX TX RF chains. Both architectures include networks of MUXs/DEMUXs intended for efficient signal routing between either the taps or AUX TXs and the transceiver antennas. 
\item We propose a general optimization framework for the joint design of analog SI cancellation and digital transceiver BF at FD MIMO nodes. 
\item We present an example algorithmic design for the analog cancellation parameters as well as the digital TX precoder and RX combiner that targets at the maximization of the FD sum rate performance. 
\item Extensive simulation results incorporating realistic models for non-ideal hardware for both proposed analog canceller architectures are presented. We compare both designed low complexity FD MIMO schemes with lately available ones in terms of hardware complexity and achievable performance.
\end{itemize}

The outline of the paper is as follows. The considered system and signal models are included in Sec$.$~\ref{sec:System_Model}, whereas Sec$.$~\ref{sec:Analog_Canceller} presents our new analog canceller architectures. Our novel general optimization framework for FD MIMO operation is provided in Sec$.$~\ref{subsec:BDC}, and Sec$.$~\ref{sec:Solution} presents an example optimization problem together with a detailed low complexity solution. Simulation results are presented and discussed in Sec$.$~\ref{sec:Results}, while Sec$.$~\ref{sec:Concl} concludes the paper and summarizes some future research directions.

\textit{Notation:} Vectors and matrices are denoted by boldface lowercase and boldface capital letters, respectively. The transpose and Hermitian transpose of $\mathbf{A}$ are denoted by $\mathbf{A}^{\rm T}$ and $\mathbf{A}^{\rm H}$, respectively, and $\det(\mathbf{A})$ is the determinant of $\mathbf{A}$, while $\mathbf{I}_{n}$ ($n\geq2$) is the $n\times n$ identity matrix and $\mathbf{0}_{m\times n}$ ($m\geq2$ and $n\geq1$) represents the $m\times n$ matrix with all zeros. $\|\mathbf{a}\|$ stands for the Euclidean norm of $\mathbf{a}$, operand $\odot$ represents the Hadamard entry-wise product, and ${\rm diag}\{\mathbf{a}\}$ denotes a square diagonal matrix with $\mathbf{a}$'s elements in its main diagonal. $[\mathbf{A}]_{i,j}$, $[\mathbf{A}]_{(i,:)}$, and $[\mathbf{A}]_{(:,j)}$ represent $\mathbf{A}$'s $(i,j)$-th element, $i$-th row, and $j$-th column, respectively, while $[\mathbf{a}]_{i}$ denotes the $i$-th element of $\mathbf{a}$. $\Real$ and $\Compl$ represent the real and complex number sets, respectively, $\mathbb{E}\{\cdot\}$ is the expectation operator, and $|\cdot|$ denotes the amplitude of a complex number.

\section{System and Signal Models}\label{sec:System_Model} 
\begin{figure}[!t]
	\begin{center}
	\includegraphics[width=0.8\textwidth]{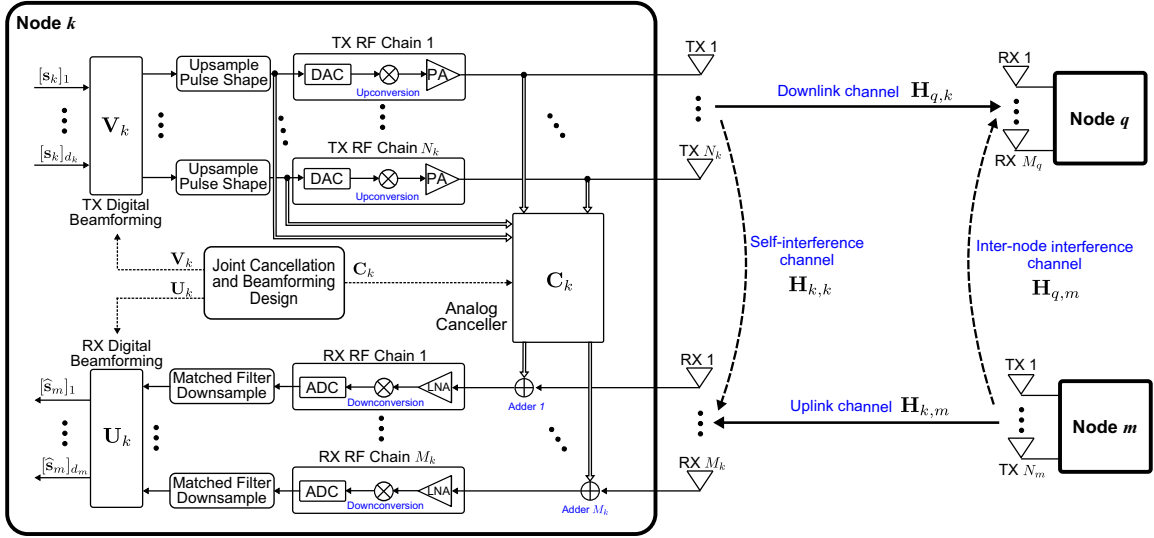}
	\caption{Schematic diagram of the considered system model and the proposed FD MIMO architectural components. The FD MIMO node $k$ communicates with the two half duplex multi-antenna nodes $q$ and $m$, the former in the downlink and the latter in the uplink communication. Node $k$ incorporates processing blocks dedicated to TX and RX digital BF, analog SI cancellation, as well as to the joint design of analog cancellation and TX/RX digital BF.}
	\label{fig:FD_MIMO}
	\end{center}
\end{figure}
We consider a wireless communication system comprising of a FD MIMO node $k$ that wishes to communicate concurrently with a multi-antenna node $q$ in the downlink and a multi-antenna node $m$ in the uplink, as shown in Fig$.$~\ref{fig:FD_MIMO}. We focus on investigating efficient FD operation at a single node, as such, we henceforth assume without loss of generality that nodes $q$ and $m$ operate in half duplex mode.

Suppose that the FD MIMO node $k$ in Fig$.$~\ref{fig:FD_MIMO} is equipped with $N_k$ TX antenna elements and $M_k$ RX antenna elements. Each antenna element is attached to a dedicated TX RF chain, similarly holds for the RX antenna elements and their respective RF chains. A TX RF chain consists of a Digital to Analog Converter (DAC), a mixer which upconverts the signal from baseband to RF, and a Power Amplifier (PA). An RX RF chain consists of a Low Noise Amplifier (LNA), a mixer which downconverts the signal from RF to baseband, and an Analog to Digital Converter (ADC). At the TX side, upsample and pulse shape processing are used to prepare the baseband signal for DAC sampling and RF transmission. At the RX side, a corresponding matched filter and downsampling is performed. The half duplex multi-antenna nodes $q$ and $m$ are assumed to have $M_q$ and $N_m$ antennas, respectively, with each antenna connected to a respective RF chain.

For presentation clarity purposes, we assume narrowband flat fading channels for our signal model. Extensions for wideband frequency selective channels are left as future work. All nodes are considered capable of performing digital BF; for simplicity, we assume hereinafter that digital TX and RX BF at the focused FD MIMO node $k$ is realized with linear filters. In particular, we assume that node $k$ makes use of the precoding matrix $\V_k\in\Compl^{N_k\times d_k}$ for processing its unit power symbol vector $\s_k\in\Compl^{d_k\times1}$ (chosen from a discrete modulation set) before transmission. The dimension of $\s_k$ satisfies $d_k\leq\min\{M_q,N_k\}$, which complies with the available spatial DoF for the downlink $M_q\times N_k$ MIMO channel. Similarly, node $m$ processes its unit power symbol vector $\s_m\in\Compl^{d_m\times1}$ (chosen again from a discrete modulation set) with a precoding matrix $\V_m\in\Compl^{N_m\times d_m}$, where $d_m\leq\min\{M_k,N_m\}$. Both the downlink and uplink transmissions are power limited according to $\mathbb{E}\{\|\V_k\s_k\|^2\}\leq {\rm P}_k$ and $\mathbb{E}\{\|\V_m\s_m\|^2\}\leq {\rm P}_m$, respectively. Following the above definitions, the baseband received signal $\y_{q}\in\Compl^{M_q\times 1}$ at node $q$ can be mathematically expressed as
\begin{equation}\label{Eq:Received_q}
\y_{q} \triangleq \H_{q,k}\V_k\s_{k} + \H_{q,m}\V_{m}\s_{m} + \n_{q}, 
\end{equation}
where $\H_{q,k}\in\Compl^{M_q\times N_k}$ is the downlink channel matrix (i$.$e$.$, between nodes $q$ and $k$), $\H_{q,m}\in\Compl^{M_q\times N_m}$ denotes the channel matrix for inter-node interference (i$.$e$.$, between nodes $q$ and $m$), and $\n_{q}\in\Compl^{M_q\times 1}$ represents the additive white Gaussian noise (AWGN) vector at node $q$ with covariance matrix $\sigma_{q}^{2}\I_{M_q}$.

Upon signal reception at the FD MIMO node $k$, analog SI cancellation is first applied to the signals received at its RX antenna elements before these signals enter to the RX RF chains, as shown in Fig$.$~\ref{fig:FD_MIMO}. Notice that the output of the analog canceller is added to the received signals before their input to the RX RF chains. We utilize the notation $\C_k\in\Compl^{M_k\times N_k}$ to represent the signal processing realized by the analog canceller. Depending on the deployed hardware components, the analog canceller can have as inputs analog or digital signals. In Sec$.$~\ref{sec:Analog_Canceller}, we will detail the hardware characteristics of our two novel analog canceller architectures. We will also show that for both architectures, the baseband representation for the output signal of the analog canceller at node $k$, which we label as $\widetilde{{\y}}_k\in\Compl^{M_k\times 1}$, is given by 
\begin{equation}\label{Eq:y_k_AC}
\widetilde{\y}_{k} \triangleq \C_k\V_k\s_{k}.
\end{equation} 
By assuming that the digitally converted and downsampled output signals of the RX RF chains at node $k$ are linearly processed in baseband by the combining matrix $\U_k\in\Compl^{d_m\times M_k}$, the estimated symbol vector $\hat{\s}_m\in\Compl^{d_m\times 1}$ for $\s_m$ is derived as  
\begin{equation}\label{Eq:Estimated_m} 
\hat{\s}_m \triangleq  \U_k\left(\y_k+\overline{\y}_k + \widetilde{{\y}}_k+\n_{k}\right), 
\end{equation}
where the complex-valued $M_k$-element vectors $\y_k$ and $\overline{\y}_k$ are the baseband representations of the received signal of interest and received SI signal, respectively, at node $k$. In addition, $\n_{k}\in\Compl^{M_k\times 1}$ denotes the received AWGN vector at node $k$ with covariance matrix $\sigma_{k}^{2}\I_{M_k}$. The vector $\y_k$ in \eqref{Eq:Estimated_m} is given by  
\begin{equation}\label{Eq:y_k_SoI}
\y_{k} \triangleq \H_{k,m}\V_{m}\s_{m},
\end{equation}
where $\H_{k,m}\in\Compl^{M_k\times N_m}$ is the uplink channel matrix (i$.$e$.$, between nodes $k$ and $m$), while $\overline{\y}_k$ is obtained as 
\begin{equation}\label{Eq:y_k_SI}
\overline{\y}_{k} \triangleq \H_{k,k}\V_{k}\s_{k},
\end{equation}
with $\H_{k,k}\in\Compl^{M_k\times N_k}$ denoting the SI channel seen at the RX antennas of node $k$ due to its own downlink transmission. 

For cases where the residual self interference in \eqref {Eq:Estimated_m} (i$.$e$.$, after performing analog cancellation and TX/RX digital BF) is above the noise floor, further digital self-interference mitigation \cite{Kor17_all} can be applied on the signal $\hat{\s}_m$ to bring the residual interference below that floor. In this paper we focus on analyzing the combined effect of analog cancellation and TX/RX digital BF, hence, we do not model a digital self-interference cancellation stage. 

\section{Novel Analog Canceller Architectures}\label{sec:Analog_Canceller} 
In this section we present the hardware components of our two novel analog SI canceller architectures. The first architecture is based on the utilization of analog taps and is thus labeled as \textit{multi-tap canceller}. The second architecture consists of AUX TXs and termed as \textit{multi-AUX-TX canceller}. The joint design of the analog canceller parameters and TX and RX digital BF will be detailed in the following sections.

\subsection{Multi-Tap Analog Canceller Architecture}\label{subsec:Analog_Canceller_Taps} 
\begin{figure}[!t]
	\begin{center}
	\includegraphics[width=0.8\textwidth]{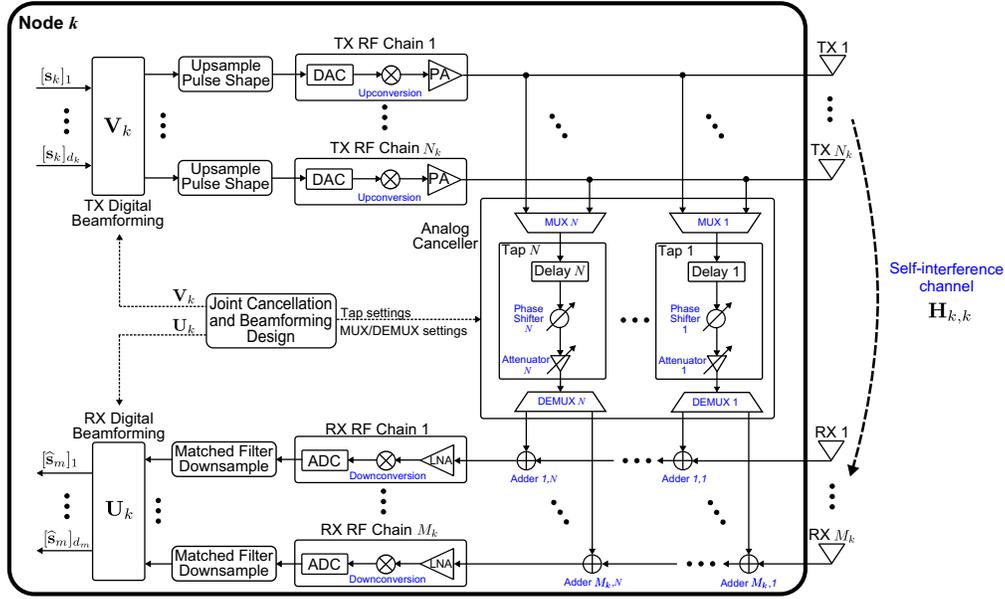}
	\caption{The proposed FD MIMO architecture at node $k$ with the multi-tap analog canceller. This canceller consists of $N$ taps, which are connected via MUXs to the outputs of the TX RF chains and via DEMUXs and adders to the inputs of the RX RF chains. With the term ``tap'' we denote a line of fixed delay, variable phase shifter, and attenuator.}
	\label{fig:FD_MIMO_ACTAPS}
	\end{center}
\end{figure}
The hardware components of the proposed multi-tap canceller for the FD MIMO node $k$ are illustrated in Fig.~\ref{fig:FD_MIMO_ACTAPS}. In this figure, $N\leq M_k N_k$ canceller taps are applied via MUXs to the outputs of the TX RF chains and via DEMUXs and adders to the inputs of the RX RF chains. One way of implementing analog RF MUXs/DEMUXs is through RF switches. With the term `tap' we denote a fixed delay-variable phase shifter-variable attenuator line, as considered in \cite{Kol16_all}. It is shown in Fig.~\ref{fig:FD_MIMO_ACTAPS} that the input of each analog canceller tap is connected to a corresponding $N_k$-to-1 MUX which allows routing of any of the $N_k$ TX RF chain signals to the input of the tap. The connection from each TX RF chain to each MUX input can be done via power dividers or directional couplers \cite{Kol16_all}. The signal that inputs to a tap undergoes a delay, phase shift, and attenuation, and this generates as an output an analog cancellation signal. The output of each tap is connected to a 1-to-$M_k$ DEMUX, which routes the cancellation signal at the output of the tap to one of the adders located just before the RX RF chains. There is a total of $M_kN$ such adders and we use ``Adder $i,j$'' to label the adder that connects DEMUX $j$ to RX RF chain $i$. Thus, the signal input to the $i$-th RX RF chain is the result of adding $N$ cancellation signals to the signal received at the $i$-th RX antenna element. Since the adders are connected to DEMUXs, some of the adders may have zero in one of the inputs depending on the DEMUXs' settings. The adders before the RX RF chains can be implemented via power combiners or directional couplers. 

As illustrated in Fig$.$~\ref{fig:FD_MIMO_ACTAPS}, analog SI cancellation is applied to the signals received at the RX antenna elements before these signals enter to the RX RF chains. Recall from Sec$.$~\ref{sec:System_Model} that we utilize the notation $\C_k\in\Compl^{M_k\times N_k}$ to represent the signal processing realized by the analog canceller. Thus, for the multi-tap canceller architecture in Fig$.$~\ref{fig:FD_MIMO_ACTAPS}, $\C_k$ captures the configuration of the MUXs/DEMUXs and the canceller tap values. We model $\C_k$ in baseband representation as the following cascade of three matrices 
\begin{equation}\label{Eq:C_k_ACTAPS}
\C_{k} \triangleq \mathbf{L}_3\mathbf{L}_2\mathbf{L}_1,
\end{equation}
where $\mathbf{L}_1\in\Real^{N \times N_k}$, $\mathbf{L}_2\in \Compl^{N\times N}$, and $\mathbf{L}_3\in\Real^{M_k\times N}$. The elements $[\mathbf{L}_1]_{i,j}$ with $i=1,2,\ldots,N$ and $j=1,2,\ldots,N_k$, and $[\mathbf{L}_3]_{i,j}$ with $i=1,2,\ldots,M_k$ and $j=1,2,\ldots,N$ take the binary values $0$ or $1$, and it must hold that 
\begin{subequations}\label{Eq:L_1_L_3}
\begin{equation}\label{Eq:L_1}
\sum_{j=1}^{N_k}[\mathbf{L}_1]_{i,j} = 1 \,\,\forall i=1,2,\ldots,N,
\end{equation}
\begin{equation}\label{Eq:L_3}
\sum_{i=1}^{M_k}[\mathbf{L}_3]_{i,j} = 1 \,\,\forall j=1,2,\ldots,N.
\end{equation}
\end{subequations}
The $i$-th row of $\mathbf{L}_1$ indicates the MUX configuration at the input of the $i$-th tap of the canceller, while the $i$-th column of $\mathbf{L}_3$ shows the DEMUX configuration at the output of the $i$-th tap of the canceller. The $\mathbf{L}_2$ in \eqref{Eq:C_k_ACTAPS} is a diagonal matrix whose complex entries represent the attenuation and phase shift of the canceller taps; particularly, the magnitude and phase of the element $[\mathbf{L}_2]_{i,i}$ with $i=1,2,\ldots,N$ specify the attenuation and phase of the $i$-th tap. Recall that the tap delays in each canceller tap are fixed and since we focus on a narrowband system, we model the effects of the $i$-th tap delay as a phase shift that is incorporated to the phase of $[\mathbf{L}_2]_{i,i}$. 

The adoption of MUXs/DEMUXs for signal routing is a novel feature of our multi-tap canceller. The flexible signal routing that is enabled by the MUXs/DEMUXs allows the use of reduced number of taps for analog cancellation, compared to the number of taps required by the designs in \cite{Bha13_all,Bha14,Kol16_all}, which require at least one tap between each TX RF chain and each RX RF chain hence at least $M_kN_k$ taps. For our proposed multi-tap canceller design, the total number of taps $N\leq M_kN_k$ is flexible and can be chosen offline as a function of node size constrains, cost per tap, or other constraints on the analog canceller hardware. Furthermore, the TX and RX digital beamformers and analog canceller will adapt to each others capabilities via our proposed joint design of analog cancellation and digital BF, which will be explained in Sec$.$~\ref{subsec:BDC}. 

\subsection{Multi-AUX-TX Analog Canceller Architecture}\label{subsec:Analog_Canceller_AUXTX} 
\begin{figure}[!t]
	\begin{center}
	\includegraphics[width=0.8\textwidth]{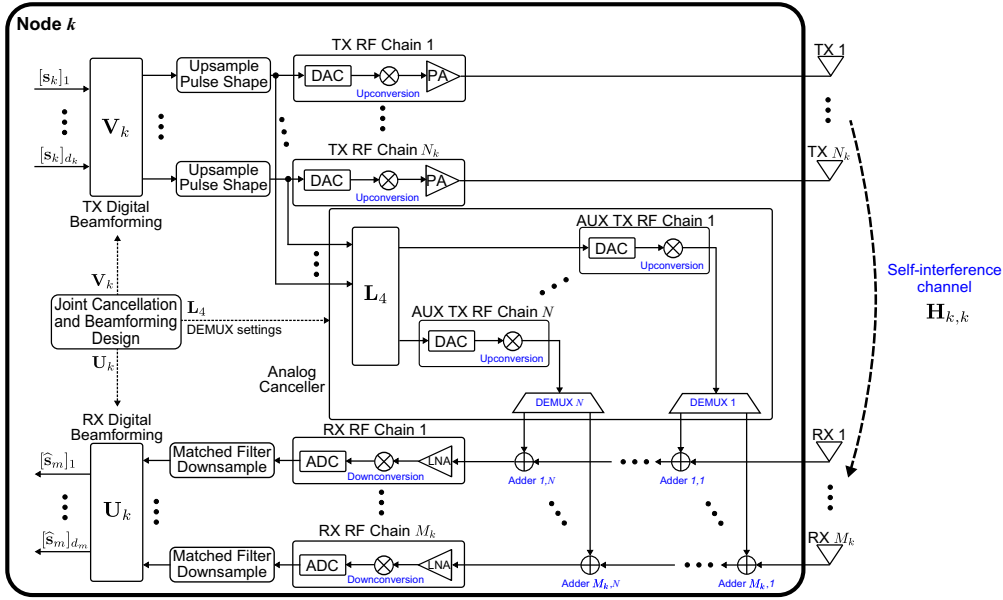}
	\caption{The proposed FD MIMO architecture at node $k$ with the multi-AUX-TX analog canceller. The analog canceller consists of $N$ AUX TX RF chains that locally generate the cancellation signal, which are connected via DEMUXs and adders to the inputs of the RX RF chains.}
	\label{fig:FD_MIMO_ACAUXTX}
	\end{center}
\end{figure}
Figure~\ref{fig:FD_MIMO_ACAUXTX} depicts the hardware components of the proposed multi-AUX-TX canceller for the FD MIMO node $k$. The analog cancellation signal is generated through $N\leq M_k$ AUX TXs, which are connected via DEMUXs and adders to the RX RF chains. An AUX TX is a TX RF chain that is used locally to generate the cancellation signal; as such, the AUX TX does not require a PA. The input to the $N$ AUX TX RF chains is generated in the digital domain and is obtained from a linear transformation of the $N_k$ output signals of the TX digital beamformer. We represent this linear transformation of the transmitted signal to generate locally the cancellation signal by the matrix $\mathbf{L}_4\in\Compl^{N\times N_k}$. It is emphasized that in the multi-AUX-TX architecture a copy of the SI signal is fed to the analog canceller in the digital domain, whereas in the multi-tap architecture depicted in Fig.~\ref{fig:FD_MIMO_ACTAPS} this connection takes place in the analog domain. However, the analog canceller outputs an analog signal for both proposed architectures. The output of each AUX TX feeds a corresponding DEMUX whose role is to route its input signal to one of the $M_k$ adders it is attached to. The latter mechanism is analogous to the DEMUX and adder connections of the multi-tap canceller described in Sec$.$~\ref{subsec:Analog_Canceller_Taps}. The baseband representation of the signal processing realized by the multi-AUX-TX canceller is modeled similar to the multi-tap case by the matrix $\C_k\in\Compl^{M_k\times N_k}$, which is now given by 
\begin{equation}\label{Eq:C_k_ACAUXTX}
\C_{k} \triangleq \mathbf{L}_5\mathbf{L}_4,
\end{equation}
where $\mathbf{L}_5\in\Real^{M_k\times N}$. The $i$-th column of $\mathbf{L}_5$ indicates the configuration of the DEMUX connected to the $i$-th AUX TX RF chain. Thus, the elements $[\mathbf{L}_5]_{i,j}$ with $i=1,2,\ldots,M_k$ and $j=1,2,\ldots,N$ take the binary values $0$ or $1$, and it must hold that 
\begin{equation}\label{Eq:L_5}
\sum_{i=1}^{M_k}[\mathbf{L}_5]_{i,j} = 1 \,\,\forall j=1,2,\ldots,N.
\end{equation}

The flexible routing of the outputs of the AUX TXs via DEMUXs that enables adjustable processing of the SI signal is a novel feature of our multi-AUX-TX canceller. The designs \cite{Hub15,Dua14_all} that adopt AUX TX RF chains do not include DEMUXs and utilize one AUX TX RF chain per RX RF chain (e$.$g$.$, $M_k$ AUX TX RF chains will be needed for node $k$ with the designs \cite{Hub15,Dua14_all}). This means that if the number of RX RF chains increases, the hardware required for the analog canceller increases as well. In contrast, our proposed multi-AUX-TX architecture can have any number $N\leq M_k$ of AUX TXs, and the effective use of the available AUX TX RF chains will be handled via the joint design of analog cancellation and digital BF, which will be detailed in the following section. 

\section{Proposed FD MIMO Optimization Framework}\label{subsec:BDC}  
In this section we present a novel FD MIMO optimization framework for the joint design of the hardware components of our analog canceller architectures described in Sec$.$~\ref{sec:Analog_Canceller} together with the TX and RX digital BF blocks included in our system model in Fig$.$~\ref{fig:FD_MIMO} in order to satisfy certain performance objectives. Capitalizing on the signal model introduced in Sec$.$~\ref{sec:System_Model}, we are particularly interested in the joint design of the analog canceller matrix $\C_k$, the digital precoding matrix $\V_k$, and the digital combining matrix $\U_k$ for the FD MIMO node $k$. We define the general objective function $f$ having as inputs the latter matrices and representing either a sole scalar performance objective, such as the average sum throughput of the FD MIMO operation, or a multi-objective performance function \cite{J:Multiobjective}, like the average sum throughput together with energy efficiency. Our general optimization framework for the joint design of $\C_k$, $\V_k$, and $\U_k$ at node $k$ is mathematically expressed by the following general optimization problem\footnote{The proposed optimization framework focuses on the joint design of the core processing blocks at the FD MIMO node $k$ for a given power budget ${\rm P}_k$, without considering the processing at nodes $q$ and $m$. A more general problem formulation for the considered system would include in the joint optimization the power allocation between downlink and uplink as well as the RX combining at node $q$ and the TX precoding of node $m$. However, in this paper, we study FD MIMO operation at node $k$ with conventional downlink and uplink control communication, and we leave the more general joint optimization that would require additional control phases for the communication of the optimized parameters as future work.}:
\begin{equation*}\label{eq:optim}
\begin{split}
  \mathcal{OP}: &\max_{\C_k,\V_k,\U_k} f\left(\C_k,\V_k,\U_k\right)
	\\& \hspace{0.6cm}\textrm{s.t.}~~\tr\{\V_k\V_k^{\rm H}\}\leq{\rm P}_k, \hspace{3.291cm}({\rm C1})
	\\& \hspace{1.26cm}               {\rm Constraints\,\,on}\,\,\C_k\,\,{\rm structure},  \hspace{1.33cm}({\rm C2}) 
	\\& \hspace{1.26cm}               g_1\left((\H_{k,k}+\C_k)\V_k\s_k\right) \leq \boldsymbol{\lambda}_{\rm A}, \hspace{1.4cm}({\rm C3})
	\\& \hspace{1.26cm}               g_2\left(\U_k(\H_{k,k}+\C_k)\V_k\s_k\right) \leq \boldsymbol{\lambda}_{\rm D}, \hspace{0.9cm}({\rm C4})
\end{split}
\end{equation*}
where constraint $({\rm C1})$ relates to the total transmit power budget at node $k$ and constraint $({\rm C2})$ refers to the hardware capabilities of the analog canceller, which impose certain limitations on the construction of $\C_k$. It follows from the discussion in Sec$.$~\ref{subsec:Analog_Canceller_Taps} that $({\rm C2})$ for the proposed multi-tap canceller architecture specifies to  
\begin{equation}\tag{C2a}
         \C_k = \mathbf{L}_3\mathbf{L}_2\mathbf{L}_1 \,\,{\rm with}\,\,\eqref{Eq:L_1},\eqref{Eq:L_3},{\rm and \,\,} [\mathbf{L}_2]_{i,j}=0\,\,{\rm for}\,\,i,j=1,2,\ldots,N\,\,{\rm with}\,\,i\neq j, 
\end{equation}
whereas for the multi-AUX-TX canceller architecture $({\rm C2})$ can be expressed using the description of Sec$.$~\ref{subsec:Analog_Canceller_AUXTX} as 
\begin{equation}\tag{C2b}
         \C_k = \mathbf{L}_5\mathbf{L}_4 \,\,{\rm with}\,\,\eqref{Eq:L_5}. 
\end{equation}
In addition, constraint $({\rm C3})$ including the general vector function $g_1:\Compl^{M_k\times1}\rightarrow\Real_+^{M_k\times1}$ sets the threshold values inside the vector $\boldsymbol{\lambda}_{\rm A}\in\Real_+^{M_k\times1}$ on functions of the instantaneous residual SI appearing at the $M_k$ RX antenna elements after analog cancellation and before the RX RF chains. Two examples of function $g_1$ are: \textit{i}) the element-wise instantaneous powers of the residual SI signals; and \textit{ii}) their summation. For the former $g_1$ example, $({\rm C3})$ results to $|[(\H_{k,k}+\C_k)\V_k\s_k]_i|^2\leq[\boldsymbol{\lambda}_{\rm A}]_i$ with $i=1,2,\ldots,M_k$, whereas for the latter example $g_1((\H_{k,k}+\C_k)\V_k\s_k)=\|(\H_{k,k}+\C_k)\V_k\s_k\|^2$ and consequently $\boldsymbol{\lambda}_{\rm A}\equiv\lambda_{\rm A}\in\Real_+$. Finally, constraint $({\rm C4})$ with the general vector function $g_2:\Compl^{d_m\times1}\rightarrow\Real_+^{d_m\times1}$ imposes the values included in the vector $\boldsymbol{\lambda}_{\rm D}\in\Real_+^{d_m\times1}$ on functions of the $d_m$ instantaneous residual SI signals obtained after applying analog cancellation and RX digital combining. Similar to $g_1$, instances of function $g_2$ are the individual instantaneous powers of the latter $d_m$ signals as well as their summation.

The main novel components of the proposed FD MIMO optimization framework in $\mathcal{OP}$ can be summarized as follows. \textit{First}, the digital TX and RX BF design takes into explicit account the available number of analog taps $N$, or number of AUX TXs $N$, of the analog SI cancellation block. Although some available BF solutions \cite{Rii11_all,Atz16_all,Gow_18_all} for FD MIMO systems consider the presence of an analog SI canceller, the details of its hardware limitations are excluded from the BF design. \textit{Second}, the proposed FD MIMO framework is the only one that explicitly considers the case where $N<\min\{M_k,N_k\}$, i$.$e$.$, the available number of analog taps, or AUX TX RF chains, may be smaller than both the numbers of TX and RX RF chains. This is an important feature for practical FD MIMO deployments, since current analog SI cancellation solutions require either very large numbers of taps, of the order of $M_kN_k$ for the architecture proposed in \cite{Bha14}, or very large number of AUX TXs, of the order of $M_k$ for the architecture presented in \cite{Hub15}. \textit{Third}, our framework has the advantage of a more optimized utilization of the spatial DoF offered by the available multiple antennas at the FD MIMO node $k$. For example, if the analog canceller consists of only $N=1$ tap, or $N=1$ AUX TX, then its cancellation capabilities are very limited, and more spatial DoF need to be devoted from the TX and RX BF blocks for meeting the thresholds $\boldsymbol{\lambda}_{\rm A}$ and $\boldsymbol{\lambda}_{\rm D}$ in $({\rm C3})$ and $({\rm C4})$. On the other extreme, if $N$ can be afforded to be large, the digital BF design may exploit the fact that a significant part of SI mitigation is handled by the analog canceller, and thus, make use of more of the available spatial DoF for improving the quality of the incoming and outgoing signals of interest. 

\section{An Example FD MIMO Design}\label{sec:Solution}
Capitalizing on the general optimization framework for the joint design of $\C_k$, $\V_k$, and $\U_k$ at the FD MIMO node $k$ described in Sec$.$~\ref{subsec:BDC}, we hereinafter present an example joint design of analog cancellation and digital BF. We assume that there is no inter-node interference between the half duplex multi-antenna nodes $q$ and $m$ due to, for example, appropriate node scheduling \cite{Italo_EW_2016_all, GA2016_all} for the FD operation of node $k$. Extensions considering this interference for the cases where it is known at either the receiving node $q$ and/or the transmitting node $k$ or unknown to both are left for future works. The latter assumption translates to setting the channel matrix between the involved nodes as $\H_{q,m}=\0_{M_q\times N_k}$. For this case, the model given by \eqref{Eq:Received_q} for the received signal at node $q$ reduces to   
\begin{equation}\label{Eq:Received_q_1}
\y_{q} = \H_{q,k}\V_k\s_{k} + \n_{q}.
\end{equation}
We rewrite the signal model \eqref{Eq:Estimated_m} that describes the estimation for $\hat{\s}_m$ at the RX of node $k$ as
\begin{equation}\label{Eq:Estimated_m_1} 
\hat{\s}_m = \U_k\left(\H_{k,m}\V_m\s_m + \widetilde{\H}_{k,k}\V_k\s_k +\n_{k}\right), 
\end{equation}
where $\widetilde{\H}_{k,k}\in\Compl^{M_k\times N_k} $ denotes the effective SI channel after performing analog cancellation, which is defined as $\widetilde{\H}_{k,k}\triangleq\H_{k,k}+ \C_k$. 

An important performance objective function $f$ for the considered system is the FD rate defined as the sum rate of the downlink and uplink communications. We therefore focus on designing $\C_k$, $\V_k$, and $\U_k$ via the solution of the following optimization problem:
\begin{equation*}\label{eq:optim1}
\begin{split}
  \mathcal{OP}1: &\max_{\C_k,\V_k,\U_k} \mathcal{R}_{\rm DL}\left(\V_k\right) + \mathcal{R}_{\rm UL}\left(\C_k,\V_k,\U_k\right)
	\\& \hspace{0.57cm}\textrm{s.t.}~~({\rm C1}),\,\,({\rm C2}),\,\,\|[\widetilde{\H}_{k,k}\V_k]_{(j,:)}\|^2\leq\lambda_{\rm A} \hspace{0.15cm} \forall j=1,2,\ldots,M_k,
	\\& \hspace{1.3cm}              \|[\U_k]_{(i,:)}\|^2=1 \hspace{0.15cm} \forall i=1,2,\ldots,d_m.
\end{split}
\end{equation*}
In the latter problem, the achievable downlink rate $\mathcal{R}_{\rm DL}$ is a function of only the digital precoding matrix $\V_k$ and is given by  
\begin{equation}\label{Eq:DL_Rate}
\mathcal{R}_{\rm DL}\left(\V_k\right) = \log_2\left(\det\left(\I_{M_q}+\sigma_q^{-2}\H_{q,k}\V_k\V_k^{\rm H}\H_{q,k}^{\rm H}\right)\right).
\end{equation} 
Note that we have assumed capacity-achieving combining at node $q$ in \eqref{Eq:DL_Rate}, like the non-linear Minimum Mean Squared Error (MMSE) successive interference canceller \cite[Chap$.$ 2]{B:Papadias_Cellular}. The uplink rate $\mathcal{R}_{\rm UL}$ in $\mathcal{OP}1$ is a function of $\V_k$, the analog canceller matrix $\C_k$, and the digital combining matrix $\U_k$, and is derived as
\begin{equation}\label{Eq:UL_Rate}
\mathcal{R}_{\rm UL}\left(\C_k,\V_k,\U_k\right) = \log_2\left(\det\left(\I_{d_m}+\sigma_k^{-2}\U_k\H_{k,m}\V_m\V_m^{\rm H}\H_{k,m}^{\rm H}\U_k^{\rm H}\Q_k^{-1}\right)\right),
\end{equation}
where $\Q_k\in\Compl^{d_m\times d_m}$ denotes the covariance matrix of the interference-plus-noise after combining at node $k$ that can be expressed as
\begin{equation}\label{Eq:Interference_Matrix}
\Q_k = \U_k\widetilde{\H}_{k,k}\V_k\V_k^{\rm H}\widetilde{\H}_{k,k}^{\rm H}\U_k^{\rm H} + \sigma_k^2\U_k\U_k^{\rm H}.
\end{equation}
Different from downlink rate in \eqref{Eq:DL_Rate}, in \eqref{Eq:UL_Rate} and \eqref{Eq:Interference_Matrix} we include the considered linear combining matrix $\U_k$ which jointly with $\V_k$ and $\C_k$ we aim to optimally design.

Note that in the formulation of $\mathcal{OP}1$ we have relaxed constraint $({\rm C3})$ concerning the instantaneous residual SI after analog cancellation that appears in the general $\mathcal{OP}$ to an average power per RX RF chain constraint, where the average is taken over all possible transmit symbol vectors. This constraint imposes that, at the input of each of the $M_k$ RX RF chains, the average power of the SI signal for all transmitted symbols within a coherent channel block cannot be larger than the threshold $\lambda_{\rm A}$. Notice also that in $\mathcal{OP}1$ we have not included a constraint similar to $({\rm C4})$ for the residual SI signal after digital combining. Instead we have only incorporated a constraint on the norm of the rows of $\U_k$. The reason for this simplification mainly lies on $\mathcal{OP}1$'s sum rate objective function. We expect that the joint design of $\C_k$, $\V_k$, and $\U_k$ optimizing the uplink rate will naturally result in keeping the average power of the residual SI signal after both analog and digital processing at an acceptable level; acceptable level is any level allowing uplink communication. Furthermore, the unity constraint on the norm of each of the rows of $\U_k$ excludes combining solutions that result in undesired amplification of the received signals (i$.$e$.$, the signals from node $m$, SI, and AWGN).

We propose to tackle $\mathcal{OP}1$ with the following two-step approach. First, as described next in Sec$.$~\ref{subsec:Solution_Ck_prelimVk}, we consider only the downlink which is usually more rate demanding than the uplink, and obtain the pairs of $\C_k$ and $\V_k$ designs optimizing the instantaneous downlink rate while meeting their respective constraints. Then, we solve for the best pair of $\C_k$ and $\V_k$ as well as the $\U_k$ design that jointly maximize the sum rate performance, as will be explained in Sec$.$~\ref{subsec:Solution_Vk_Uk}.

\subsection{Candidate Designs for $\C_k$ and $\V_k$}\label{subsec:Solution_Ck_prelimVk}
We first formulate the following downlink rate maximization problem using \eqref{Eq:DL_Rate} for the design of $\C_k$ and $\V_k$ at node $k$:  
\begin{equation*}\label{eq:optim2}
\begin{split}
  \mathcal{OP}2: &\max_{\C_k,\V_k} \mathcal{R}_{\rm DL}\left(\V_k\right)~~\textrm{s.t.}~~({\rm C1}),\,\,({\rm C2}),
	\\& \hspace{0.4cm}\|[(\H_{k,k}+\C_k)\V_k]_{(j,:)}\|^2\leq\lambda_{\rm A} \hspace{0.15cm} \forall j=1,2,\ldots,M_k.
\end{split}
\end{equation*}
To solve the latter problem we adopt an alternating optimization approach. Specifically, supposing that a realization of the analog canceller satisfying $({\rm C2})$ is given, we seek for the TX digital precoder maximizing the downlink rate, while meeting $({\rm C1})$ and the threshold $\lambda_{\rm A}$. Note that each realization of the analog canceller corresponds to a distinct MUX/DEMUX configuration. Let us assume that for $N$ taps (or $N$ AUX TXs, depending on the underlying canceller architecture) there are in total $L$ distinct realizations for the analog canceller, where $\C_k^{(\ell)}$ with $\ell=1,2,\ldots,L$ denotes the $\ell$-th canceller realization. Recall that $N$, the number of taps or AUX TXs, is decided offline upon hardware design as a function of size constraints, cost per tap and cost per AUX TX RF chain, or other hardware constraints. Examples of realizations for the analog canceller are given at the end of this section. We use the notation $\V_k^{(\ell)}$ to represent the precoder design solving $\mathcal{OP}2$ for each specific $\C_k^{(\ell)}$. The alternating optimization approach is repeated for $\C_k^{(\ell)}$ $\forall\ell$ in order to find the best pair of canceller and precoder solving $\mathcal{OP}2$. The solution for $\V_k^{(\ell)}$ given $\C_k^{(\ell)}$ is summarized in Algorithm~\ref{DL_Precoding}. The precoder is iteratively constructed as the cascade $\F_k\G_k$ with $\F_k\in\Compl^{N_k\times\alpha}$ and $\G_k\in\Compl^{\alpha\times d_k}$, where $\alpha$ is a positive integer taking the values $1\leq\alpha\leq\alpha_{\max}$ and holds that $d_k\leq\min\{M_q,\alpha\}$. In general, $\alpha_{\max}=N_k$, however, for large transmission powers and strictly small values for $\lambda_{\rm A}$ it is advisable to set $\alpha_{\max}=\min\{M_q,N_k\}$. For each value of $\alpha$ we adopt a similar approach to \cite{SofNull_2016} for the precoding design. Particularly, its $\F_k$ component aims at minimizing the impact of the residual SI MIMO channel $\widetilde{\H}_{k,k}$, whereas the goal of the $\G_k$ component is to maximize the rate of the effective downlink channel $\H_{q,k}\F_k\in\Compl^{M_q\times\alpha}$. Intuitively, parameter $\alpha$ represents the effective number of TX antennas after squeezing SI in the $N_k-\alpha$ least dominant modes of $\widetilde{\H}_{k,k}$ via the efficient use of $\F_k$. For the cases where $\H_{q,k}\F_k$ is a MIMO channel, the precoder $\G_k$ in Step 5 of Algorithm~\ref{DL_Precoding} is given by the open-loop or closed-loop precoding for this channel derived using \cite{J:Telatar_Waterfilling}, depending on whether $\H_{q,k}$ is unknown or known, respectively, at the transmit side of node $k$. In the simulation results shown later on in Sec$.$~\ref{sec:Results} we will use open-loop precoding. When $M_q=1$ and $\alpha\geq2$, $\H_{q,k}\F_k$ is a Multiple Input Single Output (MISO) channel, and if its knowledge is available at node $k$, the optimum precoding is Maximal Ratio Transmission (MRT). If $\H_{q,k}\F_k$ is a Single Input Multiple Output (SIMO) (i$.$e$.$, for $M_q\geq2$ and $\alpha=1$) or a scalar (i$.$e$.$, for $M_q=\alpha=1$) channel, $\G_k$ is a scalar set to ${\rm P}_k^{1/2}$.  

As seen from Step $15$ of Algorithm~\ref{DL_Precoding}, the $\V_k^{(\ell)}$ solving $\mathcal{OP}2$ for a specific $\C_k^{(\ell)}$ is given by $\V_{k,1}^{(\ell)}$. This notation represents the precoder corresponding to the largest value of $\alpha$ that results in meeting constraint $\lambda_{\rm A}$; recall that $\alpha$ determines $\F_k$ and $\G_k$ dimensions. We denote the maximum value of $\alpha$ for the $\C_k^{(\ell)}$ design as $\alpha^*_\ell$, and also use the notation $\V_{k,m}^{(\ell)}$ with $m=1,2,\ldots,\alpha^*_\ell$ for the $m$-th candidate precoder solution for $\mathcal{OP}2$ given $\C_k^{(\ell)}$. Although, the included iterations for solving this problem could be terminated when $\V_{k,1}^{(\ell)}$ is found, Algorithm~\ref{DL_Precoding} computes $\V_{k,m}^{(\ell)}$ $\forall m$ meeting $\mathcal{OP}2$'s threshold $\lambda_{\rm A}$ and optimizing the downlink rate for a given $\C_k^{(\ell)}$. Among those designs, the ones corresponding to lower values of $\alpha$ (i$.$e$.$, those with increasing index $m$) naturally result in larger SI mitigation. Although this behavior is desirable for maximizing the uplink rate, $\V_{k,m}^{(\ell)}$'s with larger $m$ (i$.$e$.$, obtained from lower $\alpha$) yield lower downlink rates. On the contrary, $\V_{k,1}^{(\ell)}$ maximizing the downlink rate creates the stronger SI signal contaminating the uplink. Hence, our goal with Algorithm~\ref{DL_Precoding} is to capture this trade off and obtain $\V_{k,m}^{(\ell)}$ $\forall m$ solving $\mathcal{OP}2$ for a given $\C_k^{(\ell)}$. Running this algorithm for all $L$ possible canceller realizations finally results in the joint canceller and precoder designs $\C_k^{(\ell)}$ and $\V_{k,m}^{(\ell)}$ $\forall \ell=1,2,\ldots,L$ and $\forall m=1,2,\ldots,\alpha^*_\ell$, which are feasible candidate solutions for $\mathcal{OP}2$. Those pairs will be used in Sec$.$~\ref{subsec:Solution_Vk_Uk} for obtaining the joint analog canceller and the TX/RX digital BF solution of $\mathcal{OP}1$. 
\begin{algorithm}[t!]\caption{TX Digital Precoding for a Given Analog Canceller}\label{DL_Precoding}
\begin{algorithmic}[1]
\Statex \textbf{Input:} ${\rm P}_k$, $\H_{k,k}$, $\H_{q,k}$, $\lambda_{\rm A}$, $\alpha_{\max}$, and a realization of $\C_k^{(\ell)}$ satisfying constraint $({\rm C2})$.
\State Obtain $\D_k$ including the $N_k$ right-singular vectors of $\H_{k,k}+\C_k^{(\ell)}$ corresponding to the singular values in descending order.
\State Set $m=0$.
\For{$\alpha=\alpha_{\max},\alpha_{\max}-1,\ldots,2$}
		\State Set $\F_k=[\D_k]_{(:,N_k-\alpha+1:N_k)}$.
		\State Set $\G_k$ as the optimum precoding for the effective downlink MIMO (or MISO) channel
		       \Statex \hspace{0.61cm}$\H_{q,k}\F_k$ given ${\rm P}_k$.
		\If{$\|[(\H_{k,k}+\C_k^{(\ell)})\F_k\G_k]_{(j,:)}\|^2\leq\lambda_{\rm A}$ $\forall j=1,2,\ldots,M_k$,}
		   \State Set $m=m+1$.
			 \State Store $\V_{k,m}^{(\ell)}=\F_k\G_k$ for the given $\C_k^{(\ell)}$.
		\EndIf
\EndFor
\State Set $\F_k=[\D_k]_{(:,N_k)}$ and $\G_k={\rm P}_k^{1/2}$.
\If{$|[(\H_{k,k}+\C_k^{(\ell)})\F_k\G_k]_{j}|^2\leq\lambda_{\rm A}$ $\forall j=1,2,\ldots,M_k$,}
       \State Set $m=m+1$.
			 \State Store $\V_{k,m}^{(\ell)}=\F_k\G_k$ for the given $\C_k^{(\ell)}$.
			 \State Output $\V_k^{(\ell)}=\V_{k,1}^{(\ell)}$. 
\Else
			 \State Output that $\C_k^{(\ell)}$ does not meet the residual SI constraint $\lambda_{\rm A}$.
\EndIf
\end{algorithmic}
\end{algorithm}

Algorithm~\ref{DL_Precoding} is executed at the FD MIMO node $k$ and has as inputs the MIMO channels $\H_{k,k}$ and $\H_{q,k}$ as well as a realization $\C_k^{(\ell)}$. Both $\H_{k,k}$ and $\H_{q,k}$ can be estimated through appropriately designed training processes at nodes $k$ and $q$, respectively. The latter matrix estimation can be fed back or not to node $k$ depending on whether open-loop or closed-loop MIMO operation, respectively, is adopted. We next discuss meaningful $\C_k^{(\ell)}$ realizations for both the proposed analog SI canceller architectures that provide insights on the effects of $\C_k^{(\ell)}$ choice. Note that one can also consider reducing the search of canceller realizations in $\mathcal{OP}2$ to a realization that is a deterministic function of $\H_{k,k}$ or to a desired subset of possible realizations.   

\textit{Realizations $\C_k^{(\ell)}$ for the Multi-Tap Canceller.} For a given number of taps $N$ there are in total $\binom{M_kN_k}{N}$ ways to connect the taps from the available $N_k$ TX antennas to the available $M_k$ RX antennas. This results in at most $L=\binom{M_kN_k}{N}$ possible realizations for the multi-tap canceller. Each of those refers to a different $\C_k^{(\ell)}$ matrix and corresponds to a specific placement of the $N$ tap values inside $\C_k^{(\ell)}$; its remaining elements (i$.$e$.$, $M_kN_k-N$) need to be set to zeros. One reasonable $\C_k^{(\ell)}$ intended for satisfying the SI constraint in $\mathcal{OP}2$ is to obtain $\mathbf{L}_1$, $\mathbf{L}_2$, and $\mathbf{L}_3$ such that the resulting analog canceller matrix $\C_k^{(\ell)}$ has the $N$ tap values at the same elements with the $N$ largest in amplitude elements of $\H_{k,k}$. This $\C_k^{(\ell)}$ will result in cancelling the largest SI signal components. For example, suppose that $N_k=3$, $M_k=4$, and $N=2$ and that $[\H_{k,k}]_{2,1}$ and $[\H_{k,k}]_{4,2}$ are the two largest in amplitude elements of $\H_{k,k}$. In this case, we may design $\mathbf{L}_2={\rm diag}\{[[\H_{k,k}]_{2,1} [\H_{k,k}]_{4,2}]\}$, $[\mathbf{L}_1]_{1,1}=[\mathbf{L}_1]_{2,2}=1$, and $[\mathbf{L}_3]_{2,1}=[\mathbf{L}_3]_{4,2}=1$. Other reasonable $\C_k^{(\ell)}$'s include the orderly column-by-column and row-by-row placement of the available $N$ tap values starting with the columns and rows, respectively, of $\H_{k,k}$ having the largest Euclidean norms. For example, suppose that $N_k=3$, $M_k=4$, $N=3$, and that the second RX antenna is the one most affected by SI (i$.$e$.$, the one affected by the largest SI energy). Then, having the three tap values placed at the second row of $\C_k^{(\ell)}$ will focus on reducing the SI received at the second RX antenna element. Generally, having tap values placed at the $i$-th row results in reducing SI at the $i$-th RX antenna. In the simulation results with this architecture we opt for the latter canceller design, namely the row-by-row placement of the $N$ tap values, starting with $\H_{k,k}$'s row having the largest Euclidean norm and continuing with the rest rows in descending ordering of Euclidean norms, if there are more taps to be assigned.

\textit{Realizations $\C_k^{(\ell)}$ for the Multi-AUX-TX Canceller.} To satisfy the constraint of $N$ AUX TXs, each canceller matrix needs to have $M_k-N$ all-zero rows. The $N$ nonzero rows specify the connection of the DEMUXs and the linear operation applied by $\mathbf{L}_4$. There are in total $\binom{M_k}{N}$ ways to connect the output of the $N$ AUX TXs to the $M_k$ RX antennas, and each way corresponds to a specific placement of the non-zero rows inside the canceller matrix. This results in at most $L=\binom{M_k}{N}$ possible realizations for the multi-AUX-TX canceller. One reasonable $\C_k^{(\ell)}$ realization, which we use in our simulation results for this architecture, corresponds to the case where the AUX TX RF chains are connected to the antennas that are receiving the largest SI energy. This realization targets $\H_{k,k}$'s rows having the largest Euclidean norms. Connecting the $i$-th AUX TX RF chain to the $j$-th RX antenna corresponds to setting $[\mathbf{L}_5]_{j,i}=1$.

\subsection{Joint Design of $\C_k$, $\V_k$, and $\U_k$}\label{subsec:Solution_Vk_Uk}
Using the candidate designs $\C_k^{(\ell)}$ and $\V_{k,m}^{(\ell)}$ $\forall$$\ell=1,2,\ldots,L$ and $\forall$$m=1,2,\ldots,\alpha^*_\ell$ for solving $\mathcal{OP}2$ from the approach in Sec$.$~\ref{subsec:Solution_Ck_prelimVk}, we now proceed to the final joint design of the analog canceller and TX/RX digital BF at node $k$ maximizing the instantaneous FD rate. In particular, we formulate the following optimization problem using \eqref{Eq:DL_Rate} and \eqref{Eq:UL_Rate} for the computation of the best pair of $\C_k^{(\ell)}$ and $\V_{k,m}^{(\ell)}$ together with the optimum $\U_k$:
\begin{equation*}\label{eq:optim3}
  \mathcal{OP}3:\!\!\max_{\U_k,\big\{\C_k^{(\ell)},\big \{\V_{k,m}^{(\ell)}\big\}_{m=1}^{\alpha^*_\ell}\big\}_{\ell=1}^L}\!\!{R}_{\rm DL}\left(\V_{k,m}^{(\ell)}\right) + \mathcal{R}_{\rm UL}\left(\C_k^{(\ell)},\V_{k,m}^{(\ell)},\U_k\right)~\textrm{s.t.}~\|[\U_k]_{(i,:)}\|^2=1 \hspace{0.15cm} \forall i=1,2,\ldots,d_m.
\end{equation*}

To solve $\mathcal{OP}3$ we adopt the following exhaustive search approach. For each of the $\sum_{\ell=1}^L\alpha_\ell^*$ pairs of analog canceller and TX digital precoder obtained in the previous step as candidate designs for solving $\mathcal{OP}2$, we compute $\U_k$ maximizing the uplink rate given by \eqref{Eq:UL_Rate}, while meeting its respective constraint included in both $\mathcal{OP}1$ and $\mathcal{OP}3$. Then, for each computed $\U_k$ and its corresponding $\C_k^{(\ell)}$ and $\V_{k,m}^{(\ell)}$ pair we calculate the achievable FD rate. The joint design maximizing the FD rate provides the solution for $\mathcal{OP}3$. To solve the uplink rate maximization problem we assume that $\H_{k,m}$ and $\widetilde{\H}_{k,k}$ appearing in \eqref{Eq:UL_Rate} and \eqref{Eq:Interference_Matrix} are available at node $k$ through appropriately designed training phases. With the availability of this channel knowledge and a pair of $\C_k^{(\ell)}$ and $\V_{k,m}^{(\ell)}$, it can be shown that the $\U_k$ maximizing the UL rate is given using \cite[Sec$.$ 4.2]{J:George_Elsevier_13}
by $\U_k=\boldsymbol{\Gamma} \W^{\rm H}$, where $\W\in\Compl^{M_k\times d_m}$ has as columns the $d_m$ left singular vectors of $\boldsymbol{\Lambda}_k^{-1/2}\E_k^{\rm H}\H_{k,m}$ corresponding to its respective non-zero singular values. The diagonal matrix $\boldsymbol{\Lambda}_k\in\Compl^{M_k\times M_k}$ and the matrix $\E_k\in\Compl^{M_k\times M_k}$ are obtained from the eigenvalue decomposition of the interference-plus-noise covariance matrix $\B_k\in\Compl^{M_k\times M_k}$ at node $k$, which is defined as        
\begin{equation}\label{Eq:Interference_Matrix_no_combining}
\B_k \triangleq (\H_{k,k}+\C_k^{(\ell)})\V_{k,m}^{(\ell)}[\V_{k,m}^{(\ell)}]^{\rm H}(\H_{k,k}+\C_k^{(\ell)})^{\rm H}+ \sigma_k^2\I_{M_k}.
\end{equation}
The eigenvalues of $\B_k$ are included in the main diagonal of $\boldsymbol{\Lambda}_k$, while the columns of $\E_k$ include their corresponding eigenvectors. The diagonal matrix $\boldsymbol{\Gamma} \in\Real^{d_m\times d_m}$ ensures the constraint $\|[\U_k]_{(i,:)}\|^2=1$ $\forall i=1,2,\ldots,d_m$ is met. The $i$-th entry of $\boldsymbol{\Gamma}$ is equal to $1/\|[\W^{\rm H}]_{(i,:)}\|$. For the special case of $N_m=1$ \cite{GM_ICC_2017}, which consequently results in $d_m=1$, the solution combining vector $\w_k\triangleq\W_k\in\Compl^{1\times M_k}$ simplifies to the eigenvector corresponding to the maximum eigenvalue of the matrix $\A_k\in\Compl^{M_k\times M_k}$ given by \cite{Sadek2007_TSP_all}   
\begin{equation}\label{Eq:Solution_single_stream}
\A_k \triangleq {\rm P}_m\B_k^{-1}\h_{k,m}\h_{k,m}^{\rm H},
\end{equation}
where we have used the notation $\h_{k,m}\triangleq\H_{k,m}\in\Compl^{M_k\times1}$. We note that for the practical case of imperfect analog cancellation, significant gains with the considered RX digital combining are feasible only when it holds $M_k-d_k\geq d_m$.

\subsection{Remarks}\label{subsec:Remarks}
We next provide some subtleties of our example FD MIMO design and possible extensions. We note however that, even without the following extensions, our presented design outperforms the SotA solutions, as will be shown in Sec$.$~\ref{sec:Results} including our performance evaluation results.

\textit{Remark 1:} The presented solutions of $\mathcal{OP}3$ for the analog cancellation and TX/RX digital BF are functions of the MIMO channel matrices $\H_{k,k}$, $\H_{k,m}$, and $\H_{q,k}$. This implies that the update of the BF settings as well as the settings of the canceller (values for the taps or AUX TX RF chains as well as MUX/DEMUX configurations) depend on the coherence time of the involved wireless channels.   

\textit{Remark 2:} Solving $\mathcal{OP}2$ is feasible when there exists at least one pair of $\C_k^{(\ell)}$ and $\V_{k,m}^{(\ell)}$ meeting the $\lambda_{\rm A}$ constraint. When such a pair does not exist, uplink communication is impossible to take place simultaneously with the downlink one (i$.$e$.$, FD communication for the given $N$ and $\lambda_{\rm A}$ is infeasible). We note that for our FD rate results appearing in Section~\ref{subsec:achievable_rates} we only focus on scenarios where solving $\mathcal{OP}2$ is feasible. For cases where a $\C_k^{(\ell)}$ and $\V_{k,m}^{(\ell)}$ pair satisfying $\lambda_{\rm A}$ does not exist, $\mathcal{OP}1$ can be solved via half duplex communication, and there is no need for a canceller design. In this case, the $\mathcal{OP}1$ solution is either the precoder maximizing the downlink rate or the combiner maximizing the uplink one, depending on which of the two results in the maximum half duplex rate. If we relax the SI constraint in $\mathcal{OP}1$ and $\mathcal{OP}2$ to a subset, instead of all, $M_k$ RX RF chains (i$.$e$.$, suppose that the constraint becomes $\|[\widetilde{\H}_{k,k}\V_k]_{(j,:)}\|^2\leq\lambda_{\rm A}$ $\forall j=1,2,\ldots,M_k'$ with $M_k'<M_k$), FD communication is more probable to be feasible for a given ${\rm N}_p$ and $\lambda_{\rm A}$. This happens because with this relaxation we allow uplink communication even when there exist at most $M_k-M_k'$ RX RF chains experiencing average residual SI power larger than $\lambda_{\rm A}$. However, those saturated RX RF chains should not be considered for reliable reception, hence, they should be deactivated for uplink communication via adequate antenna selection. Under this strategy, the uplink MIMO matrix is denoted by $\H_{k,m}'\in\Compl^{M_k'\times N_m}$ being a submatrix of $\H_{k,m}$, where the rows corresponding to the saturated RX RF chains have been excluded. It is finally noted that both the value for $M_k'$, and to which specific RX RF chains the $\lambda_{\rm A}$ constraint is imposed, will impact the achievable uplink rate, and hence the feasible FD communication.        

\section{Simulation Results and Discussion}\label{sec:Results}
The performance of the wireless communication scenario illustrated in Fig.~\ref{fig:FD_MIMO} using the FD MIMO design presented in Sec$.$~\ref{sec:Solution} is evaluated. In Sec$.$~\ref{subsec:Schemes} we describe the SotA solutions with which the proposed solutions will be compared. The simulation parameters and assumptions are then detailed in Sec$.$~\ref{subsec:Parameters}, whereas the SI mitigation capability and achievable rate results for different hardware complexity levels are presented in Secs$.$~\ref{subsec:residual_analysis} and \ref{subsec:achievable_rates}.

\subsection{Compared FD MIMO Designs}\label{subsec:Schemes}       
We compare our novel FD MIMO design versus the combined cancellation and spatial suppression design presented in~\cite{Rii11_all} as well as the digital BF design proposed in~\cite{SofNull_2016}. We note that the designs presented in \cite{Atz16_all, Zha12_all} were not considered in the results that follow due to the fact that they are only applicable to UpLink (UL) and DownLink (DL) communication with $d_k=d_m=1$, whereas our proposed solutions hold for $d_k,d_m\geq1$. A detailed description of the FD MIMO designs that will be compared is provided below.

\textit{Design 1: Proposed with $N$ taps.} This is our proposed FD MIMO design with a $N$-tap analog canceller. Compared with the SotA architectures \cite{Bha14, Kol16_all} requiring at least $M_kN_k$ taps, our canceller results in $100(1-N/(M_kN_k))$\% reduction in the required taps' numbers. The TX/RX digital BF as well as the settings for the canceller at the FD MIMO node $k$ are computed as presented in Sec$.$~\ref{sec:Solution}. For $\H_{q,k}\F_k$ being a MIMO channel, we have adopted open-loop MIMO precoding for the computation of $\G_k$.  

\textit{Design 2: Proposed with $N$ AUXTX.} This is our proposed FD MIMO design for the case of multi-AUX-TX canceller with $N$ AUX TX RF chains. Compared with the SotA architectures \cite{Dua14_all,Hub15} which require at least $M_k$ AUX TXs, our canceller results in $100(1-N/M_k)$\% reduction in the required number of AUX TXs. We have again used Sec$.$~\ref{sec:Solution} for the computation of TX/RX digital BF as well as the canceller settings at the FD MIMO node $k$. The $\G_k$ computation was the same as for \textit{Design 1}. 

\textit{Design 3: SotA with $M_kN_k$ taps.} This refers to a combination of time domain analog cancellation with spatial suppression as proposed in \cite{Rii11_all}. The TX beamformer is designed to minimize SI caused from this operation by using null space projection \cite{Rii11_all} for this communication side. The RX BF was proposed to be a MMSE filter in \cite{Rii11_all}, we however utilize the optimum combiner $\U_k$ obtained using \cite[Sec$.$ 4.2]{J:George_Elsevier_13}, as explained in Sec$.$~\ref{sec:Solution}. Hence we use the same combiner as in \textit{Designs 1} and \textit{2}. The time domain cancellation is a canceller that requires in total $M_kN_k$ taps (i$.$e$.$, one tap per TX-RX RF chain), as in the SotA schemes \cite{Bha14, Kol16_all}. We have made the same assumptions for the hardware capabilities of the taps for this design as in \textit{Design 1}.    

\textit{Design 4: SotA with $M_k$ AUXTX.} This design is similar to \textit{Design 3} but uses AUX TXs in place of the analog taps. It particularly combines time domain cancellation with spatial suppression \cite{Rii11_all}. The former is an analog canceller requiring a total of $M_k$ AUX TX RF chains (i$.$e$.$, one AUX TX RF chain per RX RF chain), as in the SotA schemes \cite{Dua14_all, Hub15}. In addition, the hardware capabilities of each AUX TX are considered the same with our \textit{Design 2}. TX digital BF is designed for SI minimization from the TX side, whereas RX digital BF is given by $\U_k$, as described in Sec$.$~\ref{sec:Solution}.

\textit{Design 5: SotA with $0$ taps/$0$ AUXTX.} This is the SoftNull method presented in \cite{SofNull_2016} that does not adopt analog cancellation, relying solely on TX digital BF to reduce SI at the RX antennas of node $k$. Any residual SI is handled by the RX digital combiner. The combiner $\U_k$ used in the previous designs is used for the latter purpose.

\subsection{Simulation Parameters}\label{subsec:Parameters}       
We have assumed Rayleigh fading and a path loss of $110$dB for both the DL $\H_{q,k}$ and UL $\H_{k,m}$ channels. The SI channel $\H_{k,k}$ is assumed to be subject to Ricean fading with $\kappa$-factor equal to $35$dB and path loss of $40$dB \cite{Dua12_all}. All involved wireless channels are assumed to be Independent and Identically Distributed (IID), and perfectly estimated at the receivers (i$.$e$.$, at the RXs of nodes $k$ and $q$). We have used $1000$ independent channel realizations for all statistical results. The DL transmit power ${\rm P}_k$ was set between $10$dBm and $40$dBm, and the UL transmit power ${\rm P}_m$ was set 20dB lower, hence spanning a range from $-10$dBm to $20$dBm \cite{Jos17_all}. The noise floor at node $q$ is $-90$dBm and at node $k$ is $-110$dBm. The latter values are typical ones for small cell base stations and mobile terminals. Following the findings of \cite{Sab14_all} we consider a $14$-bit ADC at node $k$ that renders digital SI mitigation of approximately $50$dB feasible. This means that for the noise floor of $-110$dBm at node $k$ the residual SI after analog cancellation (i$.$e$.$, at each RX RF chain's input) must be less than $-60$dBm. In Appendix~\ref{sec:Appendix} we detail the two realistic models used for simulating non-ideal analog canceller hardware. The one model concerns the proposed multi-tap canceller architecture and the other the multi-AUX-TX one. According to these models, the multi-tap canceller is capable of delivering approximately $60$dB of analog cancellation per tap, whereas the multi-AUX-TX canceller offers approximately $35$dB of cancellation per AUX TX RF chain. 

\subsection{Self-Interference Mitigation Capability}\label{subsec:residual_analysis}    
\begin{figure}[!t]
	\begin{center}
   \includegraphics[width=0.5\textwidth]{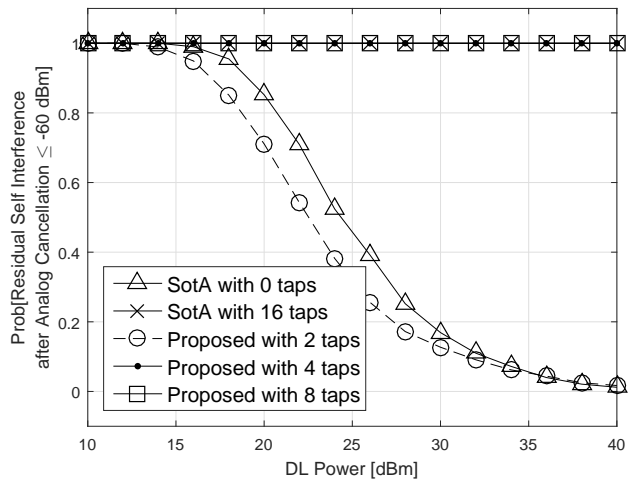}
   \caption{Probability of the residual SI power at each of the RX RF chains being less or equal to $\lambda_{\rm A}=-60$dBm versus the DL TX power ${\rm P}_k$ for the multi-tap canceller with $M_k=N_k=4$ and $M_q=N_m=1$.}
   \label{fig:Fig_probmeetconst_1DL_1UL_taps}
	\end{center}
\end{figure}

\begin{figure}[h]
	\begin{center}
        \includegraphics[width=0.5\textwidth]{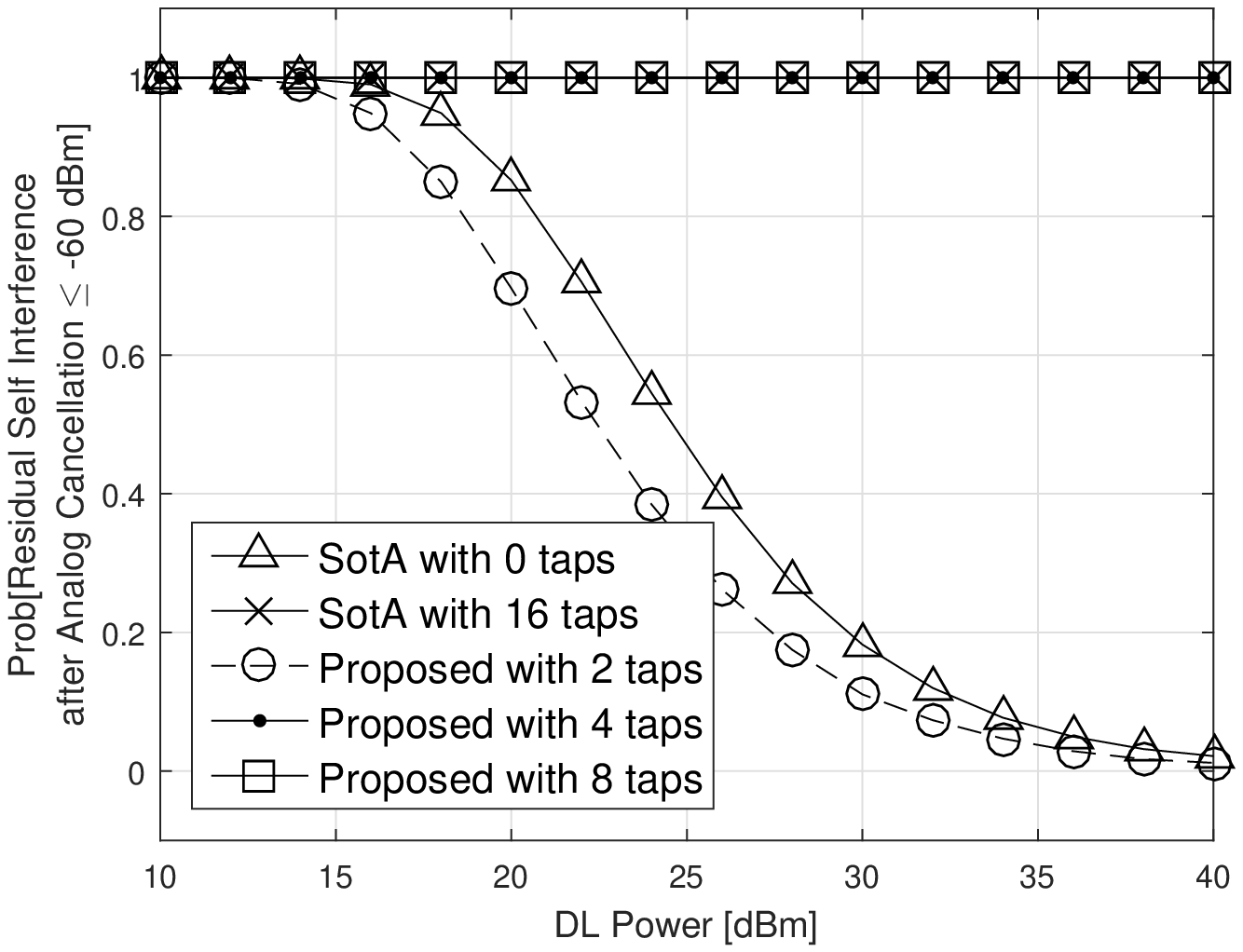}
        \caption{Probability of the residual SI power at each of the RX RF chains being less or equal to $\lambda_{\rm A}=-60$dBm versus the DL TX power ${\rm P}_k$ for the multi-tap canceller with $M_k=N_k=4$ and $M_q=N_m=4$.}
        \label{fig:Fig_probmeetconst_4DL_4UL_taps}
	\end{center}
\end{figure}

\begin{figure}[h]
	\begin{center}
        \includegraphics[width=0.5\textwidth]{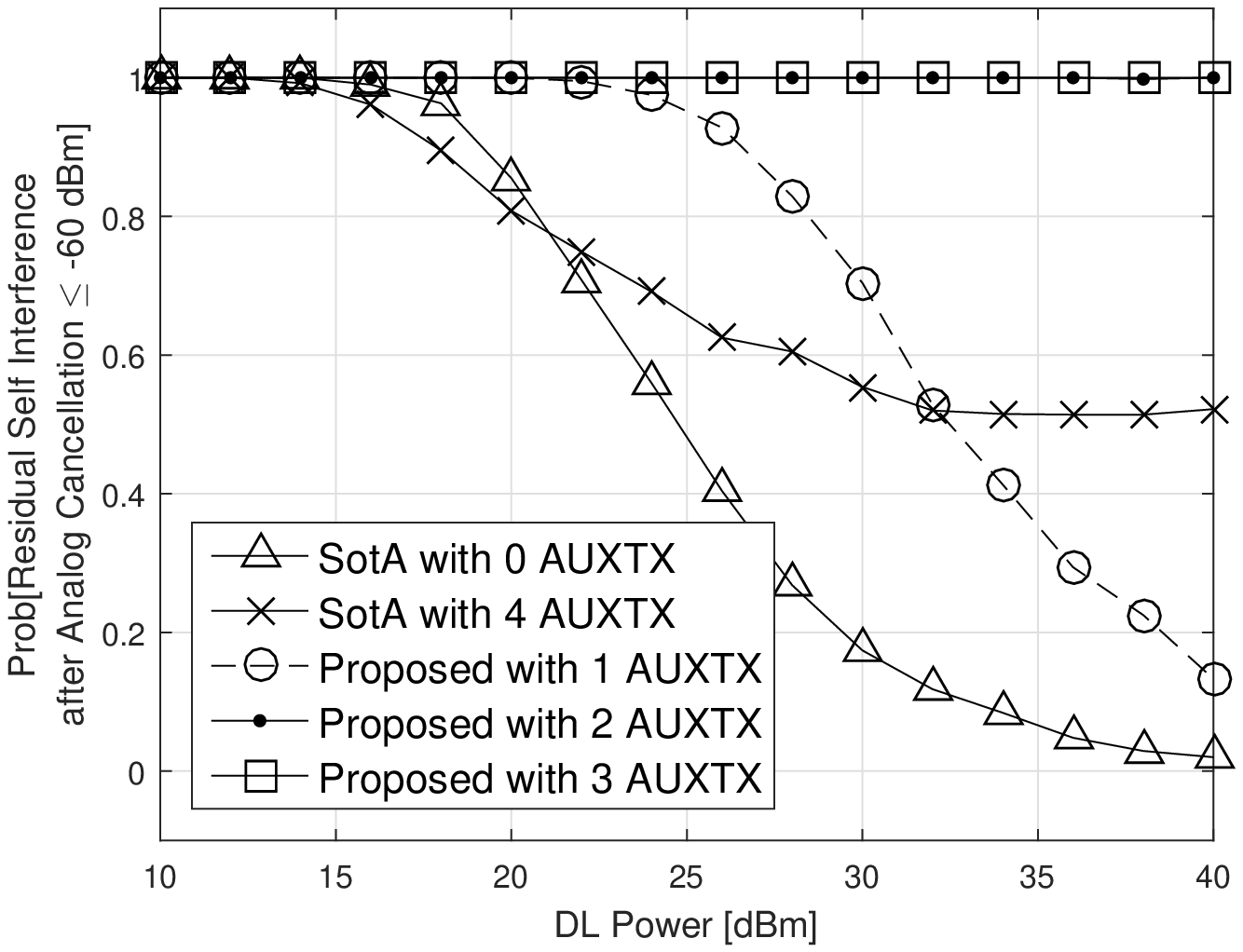}
        \caption{Probability of the residual SI power at each of the RX RF chains being less or equal to $\lambda_{\rm A}=-60$dBm versus the DL TX power ${\rm P}_k$ for the multi-AUX-TX canceller with $M_k=N_k=4$ and $M_q=N_m=1$.}
        \label{fig:Fig_probmeetconst_1DL_1UL_auxtx}
	\end{center}
\end{figure}

\begin{figure}[h]
	\begin{center}
        \includegraphics[width=0.5\textwidth]{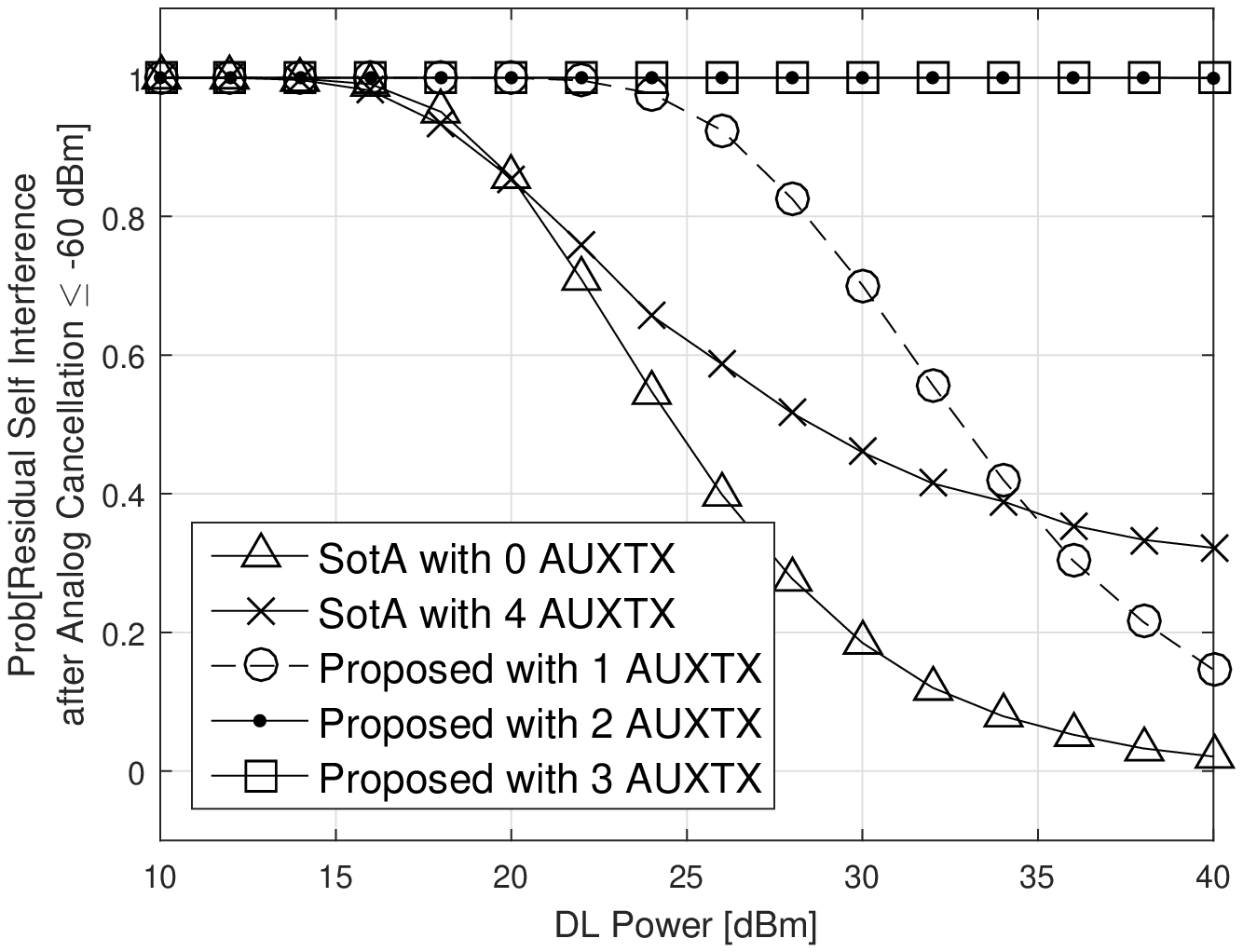}
        \caption{Probability of the residual SI power at each of the RX RF chains being less or equal to $\lambda_{\rm A}=-60$dBm versus the DL TX power ${\rm P}_k$ for the multi-AUX-TX canceller with $M_k=N_k=4$ and $M_q=N_m=4$.}
        \label{fig:Fig_probmeetconst_4DL_4UL_auxtx}
	\end{center}
\end{figure}

We consider a $4\times4$ FD MIMO node $k$ (i$.$e$.$, $M_k=N_k=4$) and two different cases for the number of antennas at nodes $q$ and $m$: the single-antenna case (i$.$e$.$, $M_q=N_m=1$) and the multi-antenna with $M_q=N_m=4$. We investigate in Figs$.$~\ref{fig:Fig_probmeetconst_1DL_1UL_taps}--\ref{fig:Fig_probmeetconst_4DL_4UL_auxtx} the probability that the residual SI after analog cancellation meets the constraint of being less than $\lambda_{\rm A} =-60$dBm. Results are shown for both proposed multi-tap and multi-AUX TX architectures for various hardware complexity levels, as implicated by different values of $N$ for the taps and AUX TXs, respectively. Within Figs$.$~\ref{fig:Fig_probmeetconst_1DL_1UL_taps}--\ref{fig:Fig_probmeetconst_4DL_4UL_auxtx} we also sketch results for SotA designs with $N=16$ taps and with $N=4$ AUX TXs, as well as for the only digital SotA solution (i$.$e$.$, $0$ taps or $0$ AUX TXs). For the latter design, we have one DL stream for the precoder, since this was the configuration yielding the largest SI reduction, however, as shown from all Figs$.$~\ref{fig:Fig_probmeetconst_1DL_1UL_taps}--\ref{fig:Fig_probmeetconst_4DL_4UL_auxtx}, for ${\rm P}_k\geq15$dBm, this design is incapable of guaranteeing residual SI power levels at any of the RX RF chains below the required $\lambda_{\rm A} =-60$dBm. Figures~\ref{fig:Fig_probmeetconst_1DL_1UL_taps} and~\ref{fig:Fig_probmeetconst_4DL_4UL_taps} demonstrate that the proposed multi-tap based design ensures that the residual SI power satisfies the $\lambda_{\rm A}$ constraint for all considered TX powers for $N=4$ and $N=8$ taps, which translates to $25$\% and $50$\% less taps compared to the SotA requiring $N=16$ taps. In addition, Figs$.$~\ref{fig:Fig_probmeetconst_1DL_1UL_auxtx} and~\ref{fig:Fig_probmeetconst_4DL_4UL_auxtx} showcase that the proposed multi-AUX TX solution with $N=2$ and $N=3$ AUX TXs is the only one based on AUX TXs that is capable of offering residual SI power below $-60$dBm for all ${\rm P}_k$ values. Actually, the SotA design with $N=4$ AUX TXs (i$.$e$.$, with $50$\% and $25$\% more AUX TXs than the $N=2$ and $N=3$ AUX TXs cases) cannot meet the residual constraint for ${\rm P}_k\geq15$dBm.  

\begin{figure}[!t]
	\begin{center}
        \includegraphics[width=0.5\textwidth]{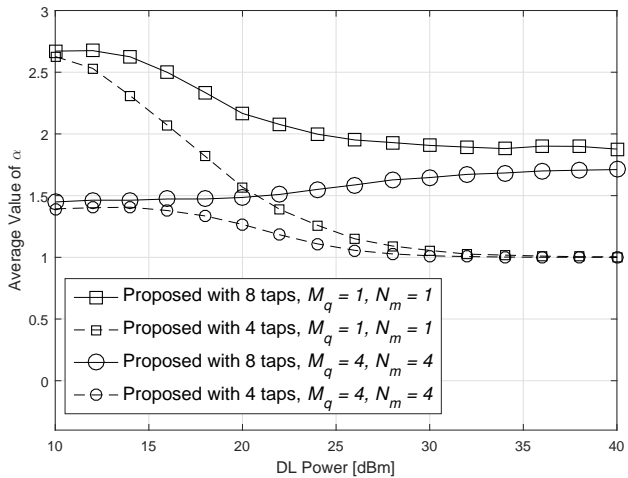}
        \caption{Average $\alpha$ value for the proposed FD MIMO design with the multi-tap architecture versus DL TX power $P_k$ and for UL TX power $P_m = P_k \mathrm{\: [dBm]} - 20 \mathrm{\: [dB]} $ for $M_k=N_k=4$.}
        \label{fig:Fig_alpha_avg_taps}
	\end{center}
\end{figure}

\begin{figure}[h]
	\begin{center}
        \includegraphics[width=0.5\textwidth]{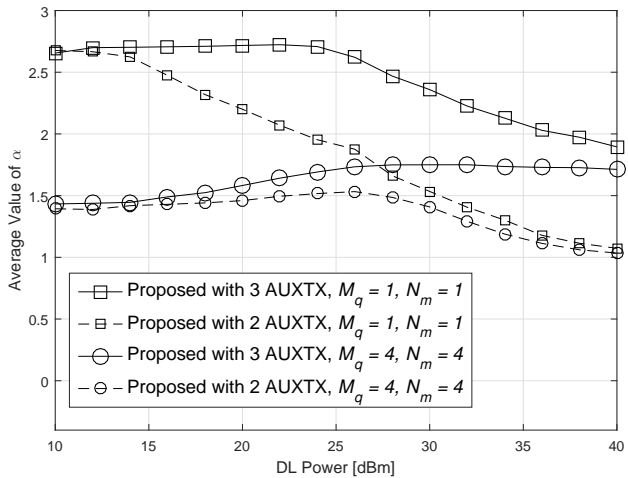}
        \caption{Average $\alpha$ value for the proposed FD MIMO design with the multi-AUX TX architecture versus DL TX power $P_k$ and for UL TX power $P_m = P_k \mathrm{\: [dBm]} - 20 \mathrm{\: [dB]} $ for $M_k=N_k=4$.}
        \label{fig:Fig_alpha_avg_auxtx}
	\end{center}
\end{figure}

We now investigate in more detail how our proposed joint analog cancellation and BF design adapts in order to meet the constraint on residual SI, while providing spatial resources for DL and UL communications. Recall that $\alpha$ used in the precoder solving $\mathcal{OP}3$ determines the effective number of TX antennas used for DL data transmission. An $\alpha$ close to $N_k$ means that the TX BF of the FD node is using more antenna resources for improving DL than for SI reduction. Therefore, $\alpha$ determines the tradeoff between acceptable SI levels as well as DL and UL achievable rates. In Figs$.$~\ref{fig:Fig_alpha_avg_taps} and~\ref{fig:Fig_alpha_avg_auxtx} we illustrate the average values of $\alpha$ chosen by our FD MIMO design as function of the DL and UL TX powers for the case of the multi-tap architecture (for $N=4$ and $N=8$ taps) and multi-AUX TX architecture (for $N=2$ and $N=3$ AUX TXs) respectively and for $M_q=N_m=1$ and $M_q=N_m=4$. From these figures we observe that for a given $M_q$ and $N_m$, the value of $\alpha$ increases as the number of taps (or AUX TXs) increases. For example, in Fig$.$~\ref{fig:Fig_alpha_avg_taps} for $M_q=N_m=1$, the values of $\alpha$ for $N=8$ taps are always larger that the values of $\alpha$ for $N=4$ taps. The more taps (or AUX TXs) the more analog canceller resources for SI mitigation, and hence less antenna resources are required for this mitigation in order to meet the residual SI constraint. This is why our algorithm chooses a larger $\alpha$ as the number of taps (or AUX TXs) increases. Thus, the results in Figs$.$~\ref{fig:Fig_alpha_avg_taps} and~\ref{fig:Fig_alpha_avg_auxtx} verify that our FD MIMO design is capable of judiciously dividing the burden of SI mitigation between the analog canceller and the TX digital BF by taking into account the resources available for analog cancellation. 

Another observation from the results in Figs$.$~\ref{fig:Fig_alpha_avg_taps} and~\ref{fig:Fig_alpha_avg_auxtx} is that as the number $M_q$ of RX antennas in DL and/or the number $N_m$ of TX antennas in UL increase, our FD MIMO design tends to be more conservative in the choice of $\alpha$ since it chooses a smaller value for $\alpha$. For example, in Fig$.$~\ref{fig:Fig_alpha_avg_taps} for the case of $N=8$ taps, the values of $\alpha$ for $M_q=N_m=1$ are larger than those for $M_q=N_m=4.$ The reason for this behavior is as follows. Recall that the number of UL streams is equal to $d_m \leq \min\{M_k,N_m\}$. Since $M_k=4$ then as $N_m$ increases from $1$ to $4$ there will be more streams in the UL communication. This increment of UL streams makes the design of TX digital BF more demanding since it has to steer SI away from these several incoming UL streams in order to maximize FD rate. Thus, our FD MIMO design chooses the small $\alpha=1$ so that the FD node can put more effort on SI mitigation. Serving less streams in DL by choosing a lower $\alpha$ allows to devote more TX spatial directions at the FD node $k$ for SI mitigation. This showcases the reconfigurability of the TX digital BF design to satisfy the FD rate maximization objective, while meeting hardware and SI constraints.

\subsection{Achievable Rates}\label{subsec:achievable_rates}  
\begin{figure}[!t]
	\begin{center}
        \includegraphics[width=0.5\textwidth]{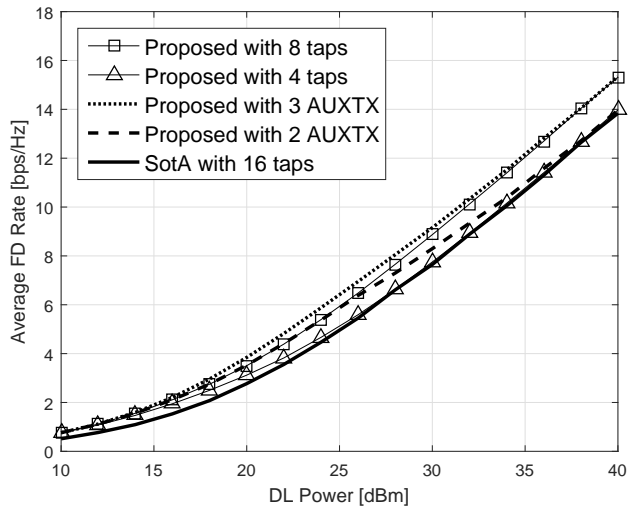}
        \caption{Average FD rates versus DL TX power $P_k$ and for UL TX power $P_m = P_k \mathrm{\: [dBm]} - 20 \mathrm{\: [dB]} $ for $M_k=N_k=4$ and $M_q=N_m=1$.}
        \label{fig:Fig_sumrates_1DL_1UL}
	\end{center}
\end{figure}

\begin{figure}[h]
	\begin{center}
        \includegraphics[width=0.5\textwidth]{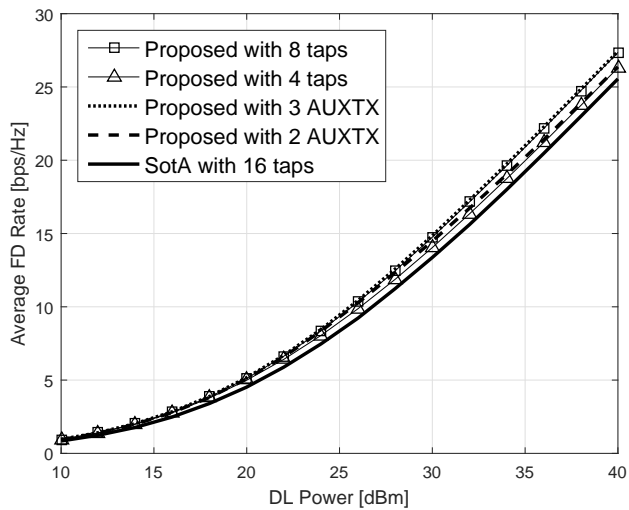}
        \caption{Average FD rates versus DL TX power $P_k$ and for UL TX power $P_m = P_k \mathrm{\: [dBm]} - 20 \mathrm{\: [dB]} $ for $M_k=N_k=4$ and $M_q=N_m=4$.}
        \label{fig:Fig_sumrates_4DL_4UL}
	\end{center}
\end{figure}

We plot the ergodic DL, UL, and FD rates in bps/Hz as functions of the TX powers for the FD MIMO systems considered in Figs$.$~\ref{fig:Fig_probmeetconst_1DL_1UL_taps}--\ref{fig:Fig_probmeetconst_4DL_4UL_auxtx} as well as in Figs$.$~\ref{fig:Fig_alpha_avg_taps} and~\ref{fig:Fig_alpha_avg_auxtx} using the algorithmic designs satisfying the constraint of having residual SI power level after analog cancellation lower than $\lambda_{\rm A}=-60$dBm. We do the same for the SotA algorithm with $N=16$ taps, which is the only design from the previous art meeting the latter SI constraint. Starting with Fig$.$~\ref{fig:Fig_sumrates_1DL_1UL} and~\ref{fig:Fig_sumrates_4DL_4UL}, the FD rate performance of the proposed multi-tap and multi-AUX TX designs with $N=\{4,8\}$ taps and $N=\{2,3\}$ AUX TXs, respectively, is illustrated in comparison with the SotA design having $N=16$ taps. As seen from both Figs$.$~\ref{fig:Fig_sumrates_1DL_1UL} and~\ref{fig:Fig_sumrates_4DL_4UL} with $M_q=N_m=1$ and $M_q=N_m=4$ respectively, the multi-tap design with $N=4$ and $N=8$ taps (i$.$e$.$, $75$\% and less $50$\% less taps than SotA) yields similar or better performance to SotA. Figures~\ref{fig:Fig_sumrates_1DL_1UL} and~\ref{fig:Fig_sumrates_4DL_4UL} also showcases the superiority of the proposed multi-AUX TX design with respect to the SotA one having $N=16$ taps. 

\begin{figure}[!t]
	\begin{center}
        \includegraphics[width=0.5\textwidth]{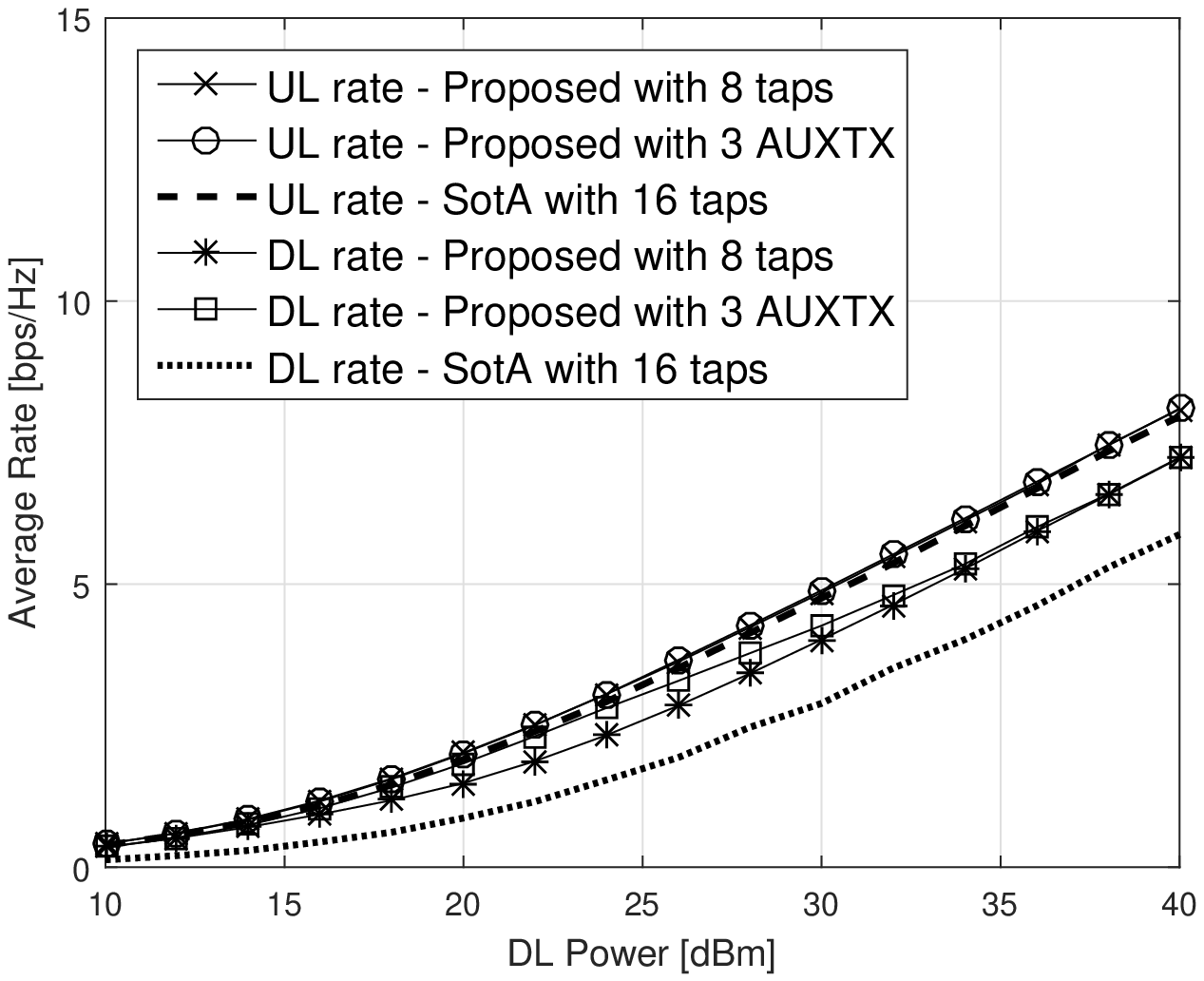}
        \caption{Average rates versus DL TX power $P_k$ and for UL TX power $P_m = P_k \mathrm{\: [dBm]} - 20 \mathrm{\: [dB]} $ for $M_k=N_k=4$ and $M_q=N_m=1$.}
        \label{fig:Fig_rates_1DL_1UL}
	\end{center}
\end{figure}

\begin{figure}[h]
	\begin{center}
        \includegraphics[width=0.5\textwidth]{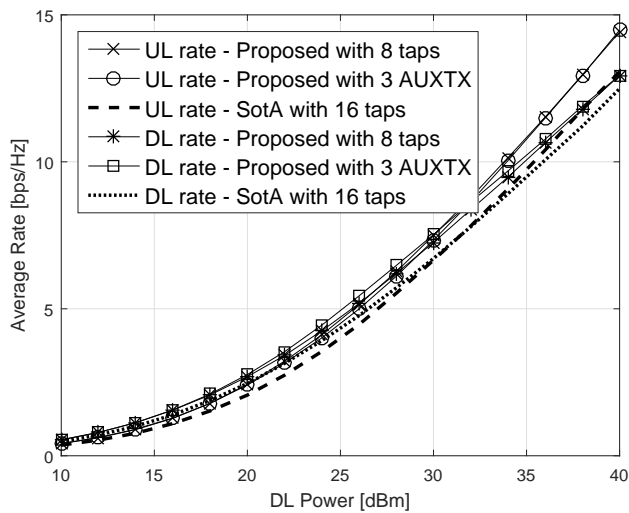}
        \caption{Average rates versus DL TX power $P_k$ and for UL TX power $P_m = P_k \mathrm{\: [dBm]} - 20 \mathrm{\: [dB]} $ for $M_k=N_k=4$ and $M_q=N_m=4$.}
        \label{fig:Fig_rates_4DL_4UL}
	\end{center}
\end{figure}

In Figs$.$~\ref{fig:Fig_rates_1DL_1UL} and~\ref{fig:Fig_rates_4DL_4UL} we focus on the achievable DL and UL rates with the proposed multi-tap and multi-AUX TX designs with $N=8$ taps and $N=3$ AUX TXs, respectively, and with the SotA design with $N=16$ taps. It is shown that our proposed TX digital precoder results in larger DL rate for $M_q=N_m=\{1,4\}$. The same trend holds for the UL rate with the proposed joint design of analog cancellation and RX digital combining. This behavior witnesses the reconfigurability of the proposed joint design, which results in both larger UL and DL gains compared to SotA as the number of antennas at nodes $q$ and $m$ increase.  

\section{Conclusion and Future Work}\label{sec:Concl}
In this paper, we have presented two novel SI mitigation schemes for FD MIMO systems with reduced hardware complexity. Each proposed scheme includes a novel analog canceller architecture, one based on analog taps and the other on AUX TXs. The main simplification of the multi-tap canceller hardware was obtained via the use of MUXs/DEMUXs for signal routing among the TX and RX RF chains and the reduced number of taps, and the joint design of the tap values and MUXs/DEMUXs configuration with the TX/RX digital BF filters. Similar simplification was gained by the multi-AUT-TX canceller hardware, where the reduced number of AUX TX RF chains were jointly designed with MUXs/DEMUXs and TX/RX digital BF. We have presented a general optimization framework for the joint design of analog SI cancellation and digital BF, and detailed a specific algorithmic solution targeting FD rate maximization. The performance evaluation results based on realistic models for non-ideal analog canceller hardware demonstrated that our proposed designs can be implemented with less cancellation elements (less taps or AUX TXs) than SotA ones, while achieving larger FD rates. For future work we intend to extend the proposed designs to wideband channels and apply the proposed framework to FD MIMO systems equipped with hybrid analog and digital transceivers \cite{Molisch_HBF}.    

\appendix
\section{Simulation models for the Analog Canceller Hardware}\label{sec:Appendix}
We next present two simulation models for non-ideal analog canceller hardware. The first model is considered for the proposed multi-tap canceller architecture and the second for the multi-AUX-TX canceller architecture.   

\subsection{Model for the Analog Taps}
In the ideal hardware case, the amplitude and phase of each of the analog taps take any desired arbitrary value. However, the settings for the attenuator and phase shifter comprising a tap take only discrete value steps when realistic hardware is considered. Consequently, we assume that each tap is set with steps of $0.02$dB for attenuation and of $0.13^{\rm o}$ for phase; these values match the step values reported in \cite{Kol16_all}. Thus, for each analog tap in our simulations, the phase setting has a random phase error uniformly distributed between $-0.065^{\rm o}$ and $0.065^{\rm o}$, and the amplitude setting has a random amplitude error uniformly distributed between $-0.01$dB and $0.01$dB. More specifically, in our simulations we do not use the ideal cancellation values given by $\C_k$, instead we use a more realistic noisy version given by $\widehat{\C}_k\in\Compl^{M_k\times N_k}$. The $N$ non-zero elements of $\widehat{\C}_k$ are the same $N$ non-zero elements of $\C_k$ but affected by a random phase and magnitude error. More specifically, for the $(i,j)$-th non-zero element of $\C$ due to the $n$-th analog tap, we compute its noisy version as
\begin{equation}\label{Eq:HW_FD_MIMO_ACTAPS}
[\widehat{\C}_{k}]_{i,j} = [\C_{k}]_{i,j} e^{j\alpha_n} 10^{\beta_n/20}\,\,{\rm for}\,\,i=1,2,\ldots,M_k\,\,{\rm and}\,\,j=1,2,\ldots,N_k,
\end{equation}
where $\alpha_n$ is uniformly distributed over the interval $[-0.065\pi/180,0.065\pi/180]$ and $\beta_n$ is uniformly distributed over the interval $[-0.01,0.01]$. In the latter expression, $\alpha_n$ and $\beta_n$ represent the phase and magnitude errors, respectively, due to the non-ideal hardware at the $n$-th tap. We also assume that $\alpha_n$ and $\beta_n$ $\forall n=1,2,\ldots,N$ are IID random variables. Applying analog cancellation with a tap exhibiting $0.065^{\rm o}$ phase and $0.01$dB magnitude errors, respectively, results in approximately $60$dB of SI cancellation. Hence, the considered multi-tap canceller architecture in our simulations is capable of delivering approximately $60$dB of analog cancellation per tap.

\subsection{Model for the AUX TX RF Chains}
One known characteristic of the canceller architecture based on AUX TX RF chains is that the SI signal used for cancellation at the RX side is obtained from the digital domain. Due to this fact, this cancellation signal does not include the inherit non-linearities of the actually transmitted SI signal; these non-linearities exist in real-world TX RF chain hardware. As has been described in \cite{Sah2013}, one of such non-linearities is the oscillator phase noise at the TX RF chains and AUX TX RF chains. This non-linearity source has been shown to be a dominant bottleneck for the performance of analog cancellers based on AUX TX RF chains. Thus, our model includes phase noise effects.

Let us denote by $\phi_i^A$ and $\phi_j^T$ with $i=1,2,\ldots,N$ and $j=1,2,\ldots,N_k$ the phase noise due to the $i$-th AUX TX RF chain and $j$-th TX RF chain, respectively. We use the matrix notation $\boldsymbol{\Phi}_k\in\Compl^{N\times N_k}$ to represent the imperfections due to phase noise. Each $(i,j)$-th element of this matrix captures the phase noise mismatch between the $i$-th AUX TX RF chain and $j$-th TX RF chain, and is expressed as 
\begin{equation}\label{Eq:Phi_ACAUXTX}
[\Phi_{k}]_{i,j} = e^{j\phi_i^A} - e^{j\phi_j^T} + 1,\,\,{\rm for}\,\,i=1,2,\ldots,N\,\,{\rm and}\,\,j=1,2,\ldots,N_k.
\end{equation}
In our simulations we do not use the ideal cancellation values given by matrix $\C_k$, we instead use a more realistic noisy version given by $\widehat{\C}_k$, which is computed as $\widehat{\C}_{k} = \mathbf{L}_5\widehat{\mathbf{L}}_4$, where the matrix $\widehat{\mathbf{L}}_4\in\Compl^{N\times N_k}$ is defined as $\widehat{\mathbf{L}}_{4} \triangleq \boldsymbol{\Phi}_{k}\odot\mathbf{L}_4$. Notice than in the ideal case of zero phase noise (i$.$e$.$, $\phi_i^A=\phi_j^T=0$), $\boldsymbol{\Phi}_{k}$ has all its entries equal to one, hence $\widehat{\mathbf{L}}_{4}$ and $\widehat{\C}_{k}$ become equal to $\mathbf{L}_{4}$ and $\C_{k}$, respectively. 
We model the phase noises as zero-mean normal random variables each with variance $\sigma_{\phi}^2$, and we set as $\sigma_{\phi}^2$ the value of the phase noise jitter $0.717^{\rm o}$, as computed in \cite{Sah2013} for the MAX2829 oscillator. Note that this value has been used in several full duplex experiments using the analog canceller architecture based on multiple AUX TXs \cite{Dua12_all,Dua14_all}. We also assume that $e^{j\phi_i^A}$ and $e^{j\phi_j^T}$ $\forall i=1,2,\ldots,N$ and $\forall j=1,2,\ldots,N_k$ are IID random variables; this implies that our model considers the case where the TX RF chains have independent oscillators. As was discussed in \cite{Sah2013}, the amount of analog cancellation for these type of phase noise is approximately $35$dB. This means that the considered multi-AUX-TX canceller architecture in our simulations is capable of delivering approximately $35$dB of analog cancellation per AUX TX RF chain.
\ifCLASSOPTIONcaptionsoff
  \newpage
\fi
\bibliographystyle{IEEEtran}
\bibliography{IEEEabrv,refs_FD_architecture}
\end{document}